%% file: 0-main.tex
\documentclass[aps,twocolumn,amsmath,amssymb,preprintnumbers,floatfix,prl,superscriptaddress,longbibliography]{revtex4-2}%{revtex4}%

\usepackage[utf8]{inputenc}
\usepackage{newtxtext}
\usepackage[upint]{newtxmath}
\usepackage{microtype}
\usepackage{textcomp}
\usepackage{eucal}
\usepackage{bm}
\usepackage{siunitx}
\usepackage{comment}
\usepackage{lipsum}
\usepackage{enumerate}
\usepackage{amsfonts}
\usepackage{amsmath}
\usepackage{amssymb}
\usepackage{color}
\usepackage{soul}

\newcommand{\pmag}{$p$M }

\newcommand{\df}[1]{\,\delta{\left(#1\right)}}
\newcommand{\RE}[1]{\,\mbox{Re}\left\{{#1}\right\}}

\usepackage{graphicx}
\usepackage[caption=false]{subfig}

\usepackage[colorlinks,allcolors=blue]{hyperref}
\usepackage[capitalize]{cleveref}

\usepackage{graphicx} 

\usepackage{bm}
\usepackage{outlines}
\usepackage{times}
\usepackage{xcolor}

\usepackage{cancel} %For strikethrough text. We can remove this later.

\definecolor{Red}{rgb}{1,0,0}
\definecolor{Blue}{rgb}{0,0,1}
\definecolor{Green}{rgb}{0,0.5,0}
\definecolor{Magenta}{rgb}{0.8,0,0.8}
\definecolor{Orange}{rgb}{1,0.64,0}

\date{June 7, 2025}
\begin{document}
\title{Coexistence of $p$-wave magnetism and superconductivity}

\author{Pavlo Sukhachov}
\affiliation{Center for Quantum Spintronics, Department of Physics, Norwegian \\ University of Science and Technology, NO-7491 Trondheim, Norway}
\email{pavlo.sukhachov@ntnu.no}
\email{pavlo.sukhachov@missouri.edu}
\affiliation{Department of Physics and Astronomy, University of Missouri, Columbia, Missouri, 65211, USA}
\affiliation{MU Materials Science and Engineering Institute,
University of Missouri, Columbia, Missouri, 65211, USA}
\author{Hans Gl{\o}ckner Giil}
\affiliation{Center for Quantum Spintronics, Department of Physics, Norwegian \\ University of Science and Technology, NO-7491 Trondheim, Norway}
\author{Bj{\o}rnulf Brekke}
\affiliation{Center for Quantum Spintronics, Department of Physics, Norwegian \\ University of Science and Technology, NO-7491 Trondheim, Norway}
\author{Jacob Linder}
%\email{jacob.linder@ntnu.no}
\affiliation{Center for Quantum Spintronics, Department of Physics, Norwegian \\ University of Science and Technology, NO-7491 Trondheim, Norway}

\begin{abstract}
The symmetry requirements for realizing unconventional compensated magnets with spin-polarized bands such as altermagnets have recently been uncovered. The most recent addition to this family of magnets is parity-odd or $p$-wave magnets. We demonstrate that $p$-wave magnets are perfectly compatible with superconductivity due to the spin polarization of their electron bands and that they induce unexpected spin transport phenomena. We first show that $p$-wave magnetism can coexist with conventional superconductivity regardless of the magnitude of the spin splitting. We then predict that $p$-wave magnets induce a charge-to-spin conversion, which can be strongly enhanced by the presence of superconductivity providing a way to probe the coexistence in experiments. Our results open an avenue for material combinations with a synergetic relation between spintronics and superconductivity.
\end{abstract}

\maketitle

\textit{Introduction.} Materials displaying magnetic order is a major topic in condensed matter physics. They have found utility in a range of existing technologies, and continue to attract attention due to their rich quantum physics.
A wealth of different magnetic states exist, from collinear
ferromagnets to complex spin textures such as skyrmions \cite{nagaosa_natnano_13, fert_natrevmat_17}.

The spatial texture of a magnetic order heavily influences the properties of electrons in magnetic materials. A prominent example of this is the recently discovered altermagnetism \cite{noda_pccp_16, ahn_prb_19, hayami_jpsj_19, yuan_prb_20, smejkal_sciadv_20, Ma-Liu-MultifunctionalAntiferromagneticMaterials-2021}, where the magnetic sublattices are connected via time-reversal and spatial rotation \cite{smejkal_prx_22a, smejkal_prx_22b}. This gives rise to a momentum-dependent spin polarization while the material as a whole has no net magnetization or stray field. Such a scenario is of high interest in the field of spintronics \cite{hirohata_jmmm_20}. Originating from the exchange interaction, the spin splitting can be very large in altermagnets compared to the splitting due to spin-orbit interactions.

The spin-dependent bands of a simple model for altermagnets have the same structure in momentum space as the Cooper pair wave function in high-temperature superconductors (SCs) \cite{keimer_nature_15}, both exhibiting a $d$-wave symmetry. In superfluid ${}^3$He \cite{leggett_rmp_75}, antisymmetric $p$-wave Cooper pairs emerge instead. Interestingly, such a superfluid state also has a magnetic equivalent; see Ref.~\cite{Jungwirth-Smejkal-Supefluid3HeAltermagnets-2024} for a recent perspective. 

In helimagnetic systems, the localized spin moments form a rotating pattern which causes a $p$-wave spin polarization to emerge \cite{martin_prb_12, chen_2dmater_22, mayo_arxiv_24} in the itinerant electron bands moving on top of this magnetic background. Such a spin polarization has also been predicted to occur in other systems~\cite{Schliemann-Loss-NonballisticSpinFieldEffectTransistor-2003, Bernevig-Zhang-ExactSUSymmetry-2006, Schliemann-Schliemann-ColloquiumPersistentSpin-2017, Hayami-Kusunose-SpontaneousAntisymmetricSpin-2020, Hayami-Kusunose-BottomupDesignSpinsplit-2020, Hayami-Hayami-MechanismAntisymmetricSpin-2022, hellenes_arxiv_23-v2}. The precise symmetry requirements for realization of $p$-wave spin-polarized bands from magnetic textures, such as $T\tau$ symmetry ($T$ being time-reversal and $\tau$ translation by a fixed number of lattice sites), were recently established \cite{hellenes_arxiv_23-v2, kudasov_prb_24} and a minimal effective model was derived~\cite{brekke_prl_24, hellenes_arxiv_23-v3}.

The minimal model shows that a $p$-wave magnet ($p$M) exhibits large tunneling magnetoresistance and spin-anisotropic transport \cite{brekke_prl_24}; a similar conclusion about the anisotropic transport was later reached in Ref.~\cite{hellenes_arxiv_23-v3}. Furthermore, $p$-wave magnetism affects the longitudinal electron transport in superconducting junctions~\cite{maedaTheoryTunnelingSpectroscopy2024, fukaya_arxiv_24} and provides a possibility of charge-to-spin conversion via, e.g., second-order spin current~\cite{brekke_prl_24}, transverse spin current in junctions with normal metal (NM)~\cite{salehi_arxiv_24}, and the nonrelativistic Edelstein effect~\cite{Chakraborty-Sinova-HighlyEfficientNonrelativistic-2024, Yu-Agterberg-OddparityMagnetismDriven-2025}; a spin-to-charge conversion in a $p$M, i.e., spin-galvanic effect, was recently studied in Ref.~\cite{Kokkeler-Bergeret-QuantumTransportTheory-2024}.

Since the band structure in a \pmag has the same structure in momentum space as the Cooper pair wave function in superfluid ${}^3$He, an intriguing question is to what extent \pmag can coexist with superconductivity and if the interplay of magnetism and superconductivity can give rise to new quantum phenomena. In this Letter, we show that the answers to both these questions are in the affirmative.

We demonstrate that unconventional electron spin splitting in \pmag is, in fact, perfectly compatible with spin-singlet superconductivity. This holds regardless of the strength of the spin-splitting, suggesting that a spontaneous coexistence of these two phases should be possible in a material with a $p$-wave magnetism. We also show that superconductivity plays a surprising role in charge-to-spin conversion: Applying an electric voltage to a junction with a \pmag generates a transverse spin current that is strongly enhanced by Andreev reflections. This enhanced conversion occurs in a $p$M-SC junction and when $p$-wave magnetism and superconductivity coexist intrinsically, suggesting a way to experimentally probe the coexistence. Our results establish another arena for material combinations with a synergetic relation between spintronics and superconductivity.

\textit{Proximity-induced superconductivity in $p$M.}  The coexistence of $p$-wave magnetism and superconductivity depends crucially on whether one considers pairing between normal-state electrons described by $c,c^\dag$ or the long-lived quasiparticles described by $\gamma,\gamma^\dag$ that are the elementary excitations of the \pmag state. The former scenario is relevant for hybrid structures of \pmag and SC, which we consider in this section. In this case, superconducting pairing takes place between normal-state electrons in a \pmag whereas magnetism is induced on top of the superconducting state via the inverse proximity effect.

\begin{figure}
    \centering
    \includegraphics[width=0.95\linewidth]{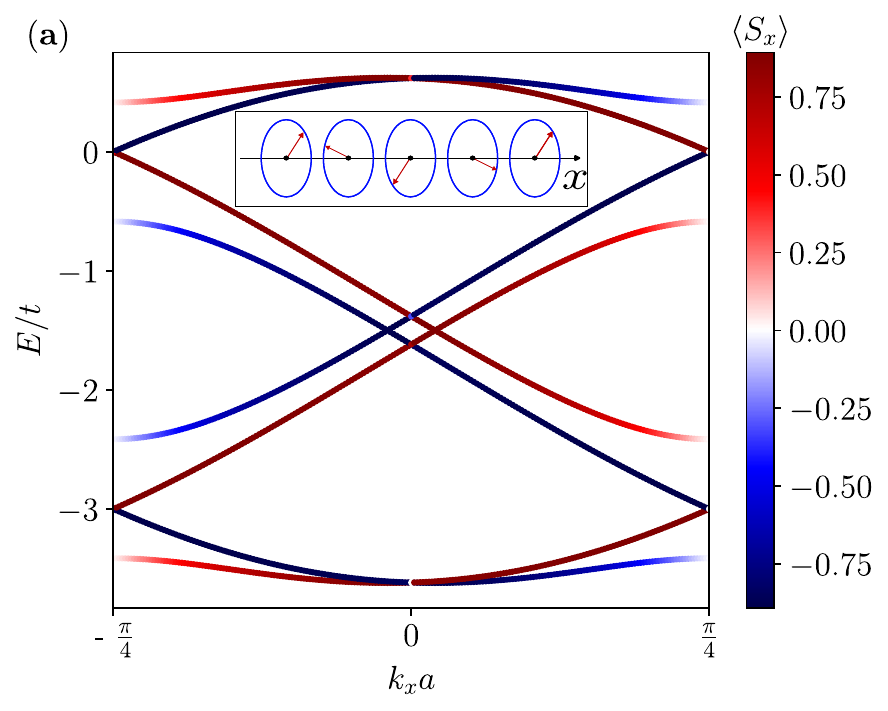}
    \\
    \includegraphics[width = 0.99 \columnwidth]{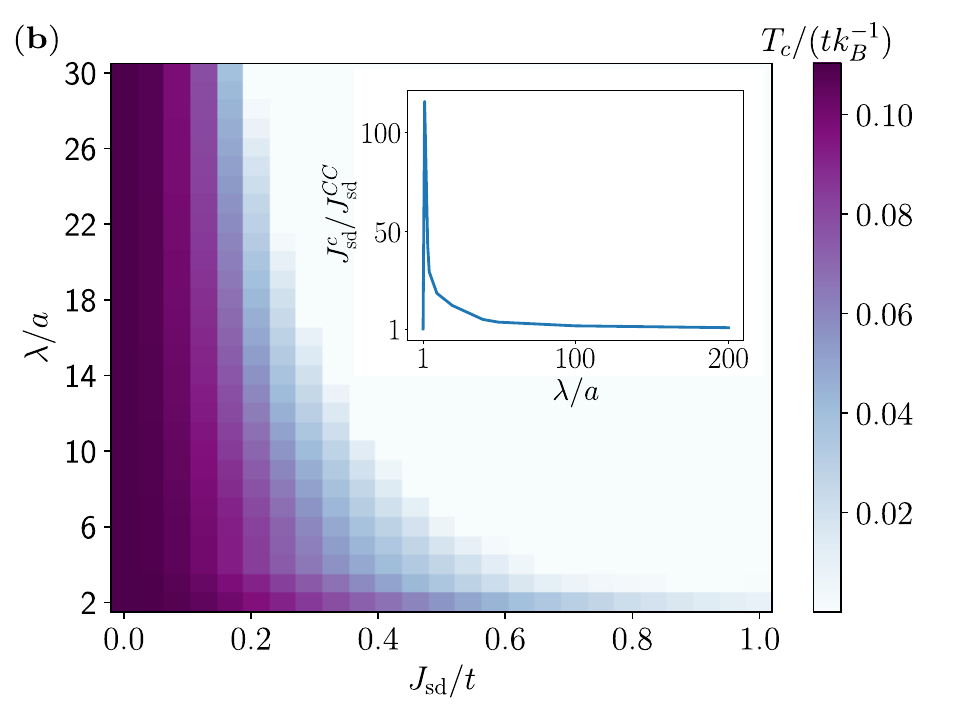}
    \caption{
    (a) The band structure for $\lambda/a = 4$, $J_\mathrm{sd} = 0.5 t$, and $\mu = -0.5t$ as a function of the momentum $k_x$ in the $x$-direction. To plot the energy spectrum, we assumed periodic boundary conditions in all directions. The insert shows the helical magnetization texture.
    (b) The critical temperature of the superconducting helimagnet as a function of the sd coupling strength $J_\mathrm{sd}$ and the rotation period $\lambda$.
    In (b), the parameters used are $\mu = -0.5t$, $N_x = 60$, $N_y = 200$, and $U = 1.7 t$. The inset shows the critical spin-splitting field $J_\mathrm{sd}^c$ at zero temperature in units of the CC critical field $J_\mathrm{sd}^{CC}$ as a function of $\lambda$. In the inset, the system length is $N_x = 200$, and we have only used values of $\lambda$ that result in zero net magnetization. The sharp peak in the inset appears because $\lambda / a = 1$ corresponds to a homogeneous ferromagnet, hence, $J_\mathrm{sd}^c =J_\mathrm{sd}^{CC}$, while $\lambda / a = 2$ is a collinear antiferromagnet, where the superconducting order is robust against sd coupling.
    }
    \label{fig:Tcoriginalbasis}
\end{figure}

To model the \pmag, we consider a lattice model with a helical magnetic texture and a mean-field on-site superconductivity described by the tight-binding Hamiltonian 
\begin{align}
\nonumber
   H =& -t\sum_{\left<i,j\right>,\sigma}  c_{i,\sigma}^{\dagger} c_{j,\sigma} 
   - J_\mathrm{sd} \sum_{i \sigma \sigma'}
    c_{i \sigma}^\dag
    [\boldsymbol S_i \cdot \bm{\sigma}]_{\sigma \sigma'}
     c_{i \sigma'}
   \\
  &
  -\mu \sum_{i, \sigma}  c_{i,\sigma}^{\dagger} c_{i,\sigma} 
  - \sum_i (\Delta_i^{\vphantom{*}} c^\dag_{i\downarrow} c^\dag_{i\uparrow} 
    + \Delta_i^* c_{i\uparrow} c_{i\downarrow}),
    \label{eq:H_tb}
\end{align}
where the spin-splitting field unit vector $\bm{S}_i$ rotates in a plane perpendicular to the helical ($x$) axis with a period $\lambda$ [see the inset in Fig.~\ref{fig:Tcoriginalbasis}(a) for a schematic of the helimagnetic chain]. The strength of the sd coupling is given by $J_\mathrm{sd}$. The superconducting potential $\Delta_i$ is calculated self-consistently, assuming a constant attractive potential $U$ in the system. The model is extended to two dimensions (2D) by making $N_y$ copies of the helical chains stacked in the $y$ direction; the number of sites in the $x$-direction is $N_x$. We apply periodic boundary conditions along the $y$-axis and open boundary conditions along the $x$ axis. The electron annihilation (creation) operator for an electron with spin projection $\sigma=\uparrow,\downarrow$ at site $i$ is denoted by $c_{i,\sigma}^{(\dagger)}$. The parameter $t$ corresponds to hopping between nearest neighbors $\left<i,j\right>$ and $\mu$ is the chemical potential. 

The band structure in the bulk of \pmag is shown in Fig.~\ref{fig:Tcoriginalbasis}(a) and demonstrates odd-parity bands with the spin polarization perpendicular to localized spins. To calculate the superconducting critical temperature in this model, we employ the lattice Bogoliubov–de Gennes formalism, assuming that the pairing takes place between the normal state electrons [see the Supplemental Material (SM) \cite{SM} for details]. As we show in Fig.~\ref{fig:Tcoriginalbasis}(b), the superconducting critical temperature decreases with an increasing period of the magnetic texture of the \pmag~\cite{champel_prb_05} and with increasing sd coupling strength $J_{\rm sd}$. This can be understood physically as follows. As the period $\lambda$ of the magnetic texture increases, the magnetization becomes more homogeneous. The case $\lambda/a=2$ ($a$ is the lattice constant) corresponds to a collinear antiferromagnet, whereas $\lambda/a \geq 3$ is a helical structure featuring $p$-wave spin-polarized bands. The $p$-wave symmetry is robust against the presence of spin-orbit coupling (SOC) for even $\lambda/a$, but is destroyed by SOC for odd $\lambda/a$ \cite{kudasov_prb_24}. However, as is shown in Ref.~\cite{hodt_arxiv_24}, deviations from a perfect $p$-wave symmetry are negligibly small in the latter case, suggesting the robustness of $p$-wave spin-polarized bands. When $\lambda$ is small or comparable to the superconducting coherence length $\xi$, the Cooper pairs experience a mostly compensated magnetization. This scenario accommodates a superconducting state coexisting with a spin-splitting that is even larger than the superconducting gap. For very large values of $\lambda/\xi$, however, the Cooper pairs experience a magnetic texture resembling a macrospin ferromagnet. Such a ferromagnet is known to destroy superconductivity at the Clogston-Chandrasekhar (CC) limit~\cite{clogston_upper_1962, chandrasekhar_note_1962}. Our numerical simulations confirm that the critical spin-splitting field is lowered toward the CC limit as $\lambda$ increases [see the inset in Fig. \ref{fig:Tcoriginalbasis}(b)]. Since the coherence length is also affected by $J_{\rm sd}$ and $\lambda$, there is no sharp boundary between the cases $\lambda\lesssim \xi$ and $\lambda \gtrsim \xi$. 
%\vfill\null

\textit{Coexistence of $p$-wave magnetism and superconductivity.} 
Having addressed a superconductor/$p$-wave magnet bilayer above, we now turn our attention to intrinsic coexistence of the two orders in the same material. To address superconductivity arising out of the normal-state of a $p$-wave magnet, we use the following low-energy 2D bulk model~\footnote{In writing Eq.~\eqref{sc-H}, we used the basis $\left\{c_{\uparrow}, c_{\downarrow}, c^{\dag}_{\uparrow}, c^{\dag}_{\downarrow}\right\}$. For this effective model, original ($c_{\sigma}$, $c_{\sigma}^{\dag}$) and diagonal ($\gamma$, $\gamma^{\dag}$) bases coincide.}:
\begin{eqnarray}
\label{sc-H}
H_{\rm BdG}(\mathbf{k}) &=& \left(\xi_k -\tilde{J}\right) \tau_z\otimes\rho_0
+\frac{\left(\mathbf{k}\cdot\bm{\zeta}\right)}{m} \tau_0\otimes\rho_z \nonumber\\
&+& \frac{\tau_x+i\tau_y}{2}\otimes \hat{\Delta}(\mathbf{k}) +\frac{\tau_x-i\tau_y}{2}\otimes \hat{\Delta}^{\dag}(\mathbf{k})
\end{eqnarray}
(see SM~\cite{SM} for the details of the model and its derivation). In essence, this is the model presented in the SM of Ref.~\cite{brekke_prl_24} in the limit of strong inter-sectoral coupling. As the model given in Eq.~\eqref{eq:H_tb}, the continuum model respects the symmetries of $p$-wave magnets, but it bears no direct connection to its tight-binding counterpart. The model is also qualitatively similar to that used in Refs.~\cite{maedaTheoryTunnelingSpectroscopy2024, salehi_arxiv_24, Ezawa-PurelyElectricalDetection-2024, fukaya_arxiv_24, ezawaThirdorderFifthorderNonlinear2024}, which resembles the electron Hamiltonian for the persistent spin helix~\cite{Schliemann-Loss-NonballisticSpinFieldEffectTransistor-2003, Bernevig-Zhang-ExactSUSymmetry-2006, Schliemann-Schliemann-ColloquiumPersistentSpin-2017}. 
The Pauli matrices $\bm{\tau}$ and $\bm{\rho}$ act in the Nambu and band spaces, respectively. Note that due to the presence of the sd coupling, the band basis is not the same as the spin basis; the difference is, however, negligible for a strong inter-sectoral coupling. To simplify the notations, we introduced $\xi_k=k^2/(2m)-\mu$, $\tilde{J}=\sqrt{J^2 +t_{\rm inter}^2}$, and $\bm{\zeta} = t_{\rm inter} \bm{\alpha}/(2\tilde{J})$. Here, $m$ is the effective mass, $t_{\rm inter}$ is the inter-sectoral coupling, $J$ is the sd coupling, and $\bm{\alpha}$ is the spin-splitting vector. The order parameter in the band basis is defined as $\hat{\Delta}(\mathbf{k}) = i\rho_y\left\{\Delta_0(\mathbf{k}) +\left[\bm{\rho}\cdot \bm{\Delta}(\mathbf{k})\right]\right\}$, where we separate the amplitude and the angular-dependent parts as $\Delta(\mathbf{k}) = \phi v(\mathbf{k})$.

As follows from the $p$-wave structure of the energy spectrum, see, e.g., the lowest two energy bands in Fig.~\ref{fig:Tcoriginalbasis}(a), a uniform spin-singlet SC order parameter is allowed in the $s$-wave channel, and the mixed-spin spin-triplet one exists in the $p$-wave channel. The equal-spin spin-triplet pairing has a nonzero center of mass momentum and will be considered elsewhere.

By using the Hamiltonian \eqref{sc-H}, we derive the gap equation via the function integral approach~\cite{SM},
\begin{equation}
\label{sc-gap}
\phi =g \frac{\nu_0}{2} \int d\xi_k \int_0^{2\pi} \frac{d\theta}{2\pi} \sum_{j} \frac{\partial \epsilon_{j,k}}{\partial \phi^{*}} \frac{1}{1+e^{-\epsilon_{j,k}/T}},
\end{equation}
where $g$ is the interaction strength, $\nu_0=m/(2\pi)$ is the normal-state DOS, $T$ is temperature, and $\epsilon_{j,k}$ are eigenvalues of $H_{\rm BdG}(\mathbf{k})$. The solution to Eq.~\eqref{sc-gap} minimizes the free energy with respect to the normal state.

We consider superconducting pairing in the $s$-wave spin-singlet and $p$-wave spin-triplet channels quantified by $\Delta_0=\phi$ and $\Delta_z(\mathbf{k})=\sqrt{2} \phi \cos{(\theta)}$, respectively. By redefining the momentum as $\mathbf{k} \to \mathbf{k}_{\sigma} = \mathbf{k} +\sigma \bm{\zeta}$ and integrating over the corresponding $\xi_{k,\sigma} \in \left[-\omega_D,\omega_D\right]$ with $\omega_D$ being the Debye frequency that provides a cutoff for the pairing interactions, we arrive at the same $s$-wave gap and the critical temperature as in the regular BCS superconductors (see, e.g., Ref.~\cite{Fossheim-Sudboe:book}). This originates from the structure of the model \eqref{sc-H}, where the pairing between fermions of opposite spins and at the opposite momenta is always possible and does not depend on the splitting of the bands. The $p$-wave mixed-spin spin-triplet pairing is also allowed but is energetically unfavorable compared to the $s$-wave gap. 

Thus, unlike superconductivity in altermagnets~\cite{Sumita-Seo-FuldeFerrellLarkinOvchinnikovStateInduced-2023, chakrabortyZerofieldFinitemomentumFieldinduced2024, Sim-Knolle-PairDensityWaves-2024, hongUnconventionalPwaveFinitemomentum2024, Bose-Paramekanti-AltermagnetismSuperconductivityMultiorbital-2024, Chakraborty-Black-Schaffer-ConstraintsSuperconductingPairing-2024}, it is possible to achieve the coexistence of magnetism and superconductivity for the simplest spin-singlet $s$-wave pairing. This result relies on the symmetry and spin polarization of the energy spectrum of $p$-wave magnets, and holds both in the low-energy model \eqref{st-H} and the full effective model with an arbitrary $t_{\rm inter}$~\cite{SM}.

\textit{Transport properties.} 
The unusual spin-polarized band structure of \pmag and the coexistence of magnetism and superconductivity are directly manifested in transport phenomena. Charge and spin transport in a non-superconducting $p$M case~\cite{brekke_prl_24, salehi_arxiv_24} and longitudinal transport in junctions with superconductors~\cite{maedaTheoryTunnelingSpectroscopy2024, fukaya_arxiv_24} have very recently been studied. To address the spin transport properties of superconducting junctions, we employ the following model:
\begin{eqnarray}
\label{st-H}
H_{\rm BdG}(x) &=& \left[\left(k_y^2-\nabla_x^2\right)/(2m) -\mu -\tilde{J}(x)\right] \tau_z\otimes\rho_0 \nonumber\\
&-& \Delta(x) \tau_y\otimes \rho_y 
+U \df{x} \tau_z\otimes \rho_0 
+\frac{\zeta_y(x)k_y}{m}\tau_0\otimes\rho_z \nonumber\\
&+& \frac{1}{2m}\left\{\zeta_x(x), -i\nabla_x\right\} \tau_0\otimes\rho_z
\end{eqnarray}
[see also Eq.~\eqref{sc-H} for the bulk Hamiltonian]. We assume the spin-singlet $s$-wave superconducting gap at $x>0$ with $\Delta(x) = \Delta \Theta{(x)}$. The spin-splitting parameter $\bm{\zeta}(x)$ and the sd coupling term $J(x)$ are nonzero either at $x<0$ in the $p$M-SC junction or at $x>0$ in the NM-superconducting $p$-wave magnet (NM-SC$p$M) one. To model a nonideal interface, we include the spin-independent barrier potential $U \df{x}$.

In calculating the transport properties of the junctions, we use the standard formalism based on the impinging, scattered, and transmitted waves (see SM~\cite{SM} for details). We focus on the transport coefficients averaged over the coordinate $x$. In this case, the contribution from the interface-localized modes \cite{linder_prl_11} becomes negligible. The transverse spin conductance is determined by integrating the corresponding spin current over all transverse momenta $k_y$
\begin{equation}
\label{st-GS-def}
G_{j,S}(V) = -e \int_{-\infty}^{\infty}\frac{dk_y}{(2\pi)^2} \left|\frac{\partial k_x}{\partial \epsilon}\right| I_{j,S}(eV,k_y),
\end{equation}
where the spin current per $k_y$ is
\begin{equation}
\label{st-JS-def}
I_{j,S}(\epsilon,k_y) = \sum_{\sigma=\pm} \RE{\Psi_{\sigma}^{\dag} \hat{v}_j \hat{s}_z \Psi_{\sigma}},
\end{equation}
with $\hat{\mathbf{v}}$ being the velocity operator, $\hat{s}_z$ being the spin operator, and $\Psi_\sigma$ being the scattering state with the spin projection $\sigma$. For the transverse spin current in a $p$-wave magnet, we have $\hat{v}_y \hat{s}_z = t_{\rm inter}\left( k_y \tau_0\otimes\rho_z +\zeta_y \tau_z\otimes\rho_0 \right)/(m\tilde{J})$. Then, the spin current reads
\begin{equation}
\label{st-JS-pM}
I_{y,S}(\epsilon,k_y) = \frac{t_{\rm inter}}{m\tilde{J}} \sum_{\sigma=\pm} \sigma k_{y,\sigma} \left[1- |a_{\sigma}(k_y)|^2 +|b_{\sigma}(k_y)|^2\right],
\end{equation}
where $a_{\sigma}(k_y)$ and $b_{\sigma}(k_y)$ are Andreev and normal reflection amplitudes, respectively, that are found by matching the wave functions and their derivatives~\footnote{In matching the derivatives, one needs to take into account the coordinate-dependence of $\zeta_x(x)$.} at $x=0$. In the NM, one has to replace $\tilde{J} \to t_{\rm inter}$ and $k_{y,s}\to k_{y}$.  Each of the terms in the square brackets contributes only when the corresponding quasiparticles, i.e., retroreflected holes and reflected electrons, are propagating. Impinging electrons are always propagating with the real wavevector $k_x=- \sigma \zeta_x + \sqrt{2m\left(\mu+\tilde{J} +\epsilon\right) -k_{y,\sigma}^2+|\bm{\zeta}|^2}$. 

\begin{figure}[ht!]
\centering
\begin{subfloat}{\includegraphics[width=0.95\columnwidth]{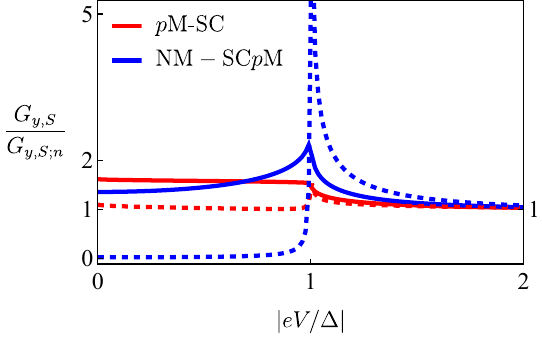}}
\end{subfloat}
\caption{\label{fig:st-G}
Transverse spin conductance in the non-superconducting parts of the $p$M-SC (red lines) and NM-SC$p$M (blue lines) junctions for an ideal contact (solid lines, $Z=0$) and in the tunneling limit (dashed lines, $Z=10$). The dependency on $\bm{\zeta}$ and other values of $Z$ can be found in SM~\cite{SM}. We normalize by the normal-state transverse spin conductance denoted as $G_{y,S;n}$ (we set $|eV|/\Delta=3$ in $G_{y,S}$) and fixed $\zeta_x=0$, $\zeta_y=0.5\,k_F$, and $J=0$. In addition, we denote $Z=U\sqrt{2m/\mu}$ and $k_F=\sqrt{2m\mu}$.
}
\end{figure}

The normalized transverse spin conductance $G_{y,S}$ is shown in Fig.~\ref{fig:st-G}. It relies on the spin-splitting vector having a nonzero component along the junction; $G_{y,S}$ grows with the spin-splitting $\zeta_y$. Due to the symmetry of the spin-polarized bands, there is no longitudinal spin current. The superconducting gap strongly affects the spin conductance, allowing for the enhanced subgap conductance in an ideal contact (see the solid lines in Fig.~\ref{fig:st-G}), and a well-pronounced peak at $|eV|\gtrsim \Delta$ in a tunneling regime (see the dashed lines in Fig.~\ref{fig:st-G}). The Andreev-enhanced transverse spin conductance is in drastic contrast with the longitudinal spin conductance in ferromagnet-SC junctions~\cite{kashiwaya_prb_99}, where both normal and Andreev reflections reduce the conductance. Whereas efficient charge-to-spin conversion is thus possible by a longitudinal charge current converting into a transverse spin current, superconductivity strongly enhances the charge-to-spin conversion rate in both $p$M-SC and NM-SC$p$M junctions. This enhancement occurs in the experimentally relevant regime of low interface transparency, such that the effect does not rely on an idealized, perfect interface.

\begin{figure}[ht!]
\centering
\begin{subfloat}{\includegraphics[width=0.8\columnwidth]{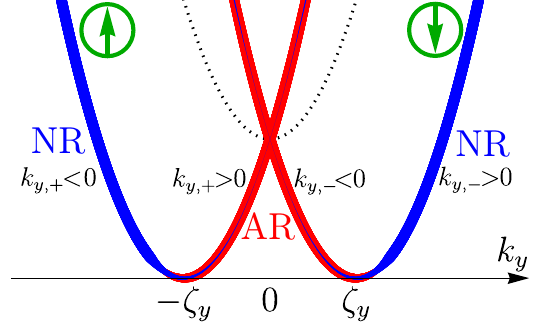}}
\end{subfloat}
\caption{\label{fig:st-transport}
Schematic depiction of spin-down (right parabola) and spin-up (left parabola) bands on the $p$M-side of the $p$M-SC junction. The bands are separated by $2\zeta_y$ along $k_y$. The dotted line corresponds to the normal-state Fermi surface on the SC side of the junction. Andreev reflection (AR) and normal reflection (NR) coefficients are shown in red and blue colors, respectively. The former relies on the overlap between the Fermi surfaces on both sides of the junction.
}
\end{figure}

The physical origin of the Andreev-enhanced spin conductance seen in Fig.~\ref{fig:st-G} is related to the momentum-space structure of the transport coefficients, which, in turn, depends on the overlap of the Fermi surfaces (see Fig.~\ref{fig:st-transport}). The coefficient $|a_s(k_y)|^2$ that determines the Andreev reflections requires an overlap between the normal-state Fermi surface in the SC and the Fermi surfaces in the \pmag (see the red color in Fig.~\ref{fig:st-transport}). This means that $s\sum_{k_y} k_{y,s} |a_s(k_y)|^2 >0$. On the other hand, the coefficient for normal reflections $|b_s(k_y)|^2$ is dominated by the parts of the Fermi surfaces in \pmag that have no overlap with that on the SC side (see the blue color in Fig.~\ref{fig:st-transport}). This results in $s\sum_{k_y} k_{y,s} |b_s(k_y)|^2<0$. Thus, since the first term in the square brackets in Eq.~\eqref{st-JS-pM} vanishes after integrating over momenta, both Andreev and normal reflections enhance the transverse spin conductance.

\textit{Concluding remarks.} 
We showed that magnetic order in $p$-wave magnets coexists with spin-singlet superconductivity. In proximity setups, coexistence quantified by the critical temperature is controlled by the period of the helical magnetic texture in the magnet and the coherence length of the superconductor. The CC limit is exceeded by almost an order of magnitude when the period of the texture $\lambda$ becomes smaller than the coherence length $\xi$ (see Fig.~\ref{fig:Tcoriginalbasis}), and approaches the CC limit at $\xi \ll \lambda$. The results for the intrinsic spin-singlet pairing in the effective low-energy model also support the coexistence of superconductivity and magnetism in a bulk $p$M.

The interplay of $p$-wave spin polarization and superconductivity is manifested in the transport properties of junctions exemplified by the transverse spin current. Contrary to the longitudinal spin conductance in ferromagnet-superconductor junctions, $p$-wave magnets allow for transverse spin conductance that is enhanced by Andreev reflections (see Fig.~\ref{fig:st-G}). This enhancement is directly related to the structure and spin polarization of the $p$-wave spin-polarized bands. A qualitatively similar normalized spin conductance is observed for $p$M-SC and NM-SC$p$M junctions, see Fig.~\ref{fig:st-G}, allowing one to identify SC$p$M in transport measurements.

The proposed coexistence and spin transport could be realized in $p$M-candidates CeNiAsO~\cite{hellenes_arxiv_23-v3}.
Helimagnets with $p$-wave polarization such as MnP, FeP, CrAs~\cite{Kallel-Bertaut-HelimagnetismMnPtypeCompounds-1974}, MnAu$_2$~\cite{Herpin-Meriel-EtudeAntiferromagnetismeHelicoidal-1961}, MnGe~\cite{Kanazawa-Tokura-LargeTopologicalHall-2011}, MnSi~\cite{Ishikawa-Roth-HelicalSpinStructure-1976}, and $\alpha$-EuP$_3$~\cite{Mayo-Ishiwata-MagneticGenerationSwitching-2022} should also allow for some of the proposed effects.

Our results provide guidance on how the interplay between $p$-wave magnetism and superconductivity, including their coexistence, can be probed and suggest a promising platform for investigating the interplay of superconductivity and magnetization thus enriching the field of spintronics. 

\textit{Acknowledgments.}  
We thank Hendrik Bentmann for pointing out the similarity of the band structure of $p$-wave magnets and the persistent spin helix structures and Erik W. Hodt for many useful discussions. This work was supported by the Research Council of Norway through Grant No. 323766 and its Centres of Excellence funding scheme Grant No. 262633 “QuSpin.” Support from Sigma2 - the National Infrastructure for High Performance Computing and Data Storage in Norway, Project No. NN9577K, is acknowledged.

\bibliography{p-wave_magnetism-restored}

\end{document}

%\date{\today}

\title{Supplemental Material\\ 
Coexistence of $p$-wave Magnetism and Superconductivity}
% \author{hss}
%\affiliation{Center for Quantum Spintronics, Department of Physics, Norwegian \\ University of Science and Technology, NO-7491 Trondheim, Norway}

\author{Pavlo Sukhachov}
\affiliation{Center for Quantum Spintronics, Department of Physics, Norwegian \\ University of Science and Technology, NO-7491 Trondheim, Norway}
\email{pavlo.sukhachov@ntnu.no}
\email{pavlo.sukhachov@missouri.edu}
\affiliation{Department of Physics and Astronomy, University of Missouri, Columbia, Missouri, 65211, USA}
\affiliation{MU Materials Science \& Engineering Institute,
University of Missouri, Columbia, Missouri, 65211, USA}
\author{Hans Gl{\o}ckner Giil}
\affiliation{Center for Quantum Spintronics, Department of Physics, Norwegian \\ University of Science and Technology, NO-7491 Trondheim, Norway}
\author{Bj{\o}rnulf Brekke}
\affiliation{Center for Quantum Spintronics, Department of Physics, Norwegian \\ University of Science and Technology, NO-7491 Trondheim, Norway}
\author{Jacob Linder}
%\email{jacob.linder@ntnu.no}
\affiliation{Center for Quantum Spintronics, Department of Physics, Norwegian \\ University of Science and Technology, NO-7491 Trondheim, Norway}

\maketitle
\tableofcontents

\section{Proximity-induced superconductivity in lattice model}
\label{sec:lattice_coex}
\input{Supplemental-pMSC-1}

\section{Intrinsic superconductivity in low-energy model}
\label{sec:analytics-coexistence}
\input{Supplemental-pMSC-2}

\section{Transport properties}
\label{sec:transport}
\input{Supplemental-pMSC-3}

\bibliography{p-wave_magnetism-restored}

%% file: Supplemental-pMSC-1.tex
In this section, we discuss the details of the lattice model for $p$-wave magnets ($p$M) and present additional results for the band structure and proximity effect. 

The Hamiltonian of the helimagnetic one-dimensional (1D) \pmag chain is
\begin{align}
\label{sm1-model-h-def}
    H = - \mu \sum_i c_{i \sigma}^\dag c_{i \sigma} - t \sum_{\langle i, j \rangle} c_{i \sigma}^\dag c_{j \sigma} - J_\mathrm{sd} \sum_{i \sigma \sigma'}
    c_{i \sigma}^\dag
    [\vec S_i \cdot \bm{\sigma}]_{\sigma \sigma'}
     c_{i \sigma'},
\end{align}
and the local spin-splitting field at site $i$ is given by the unit vector $\vec S_i$, and $J_\mathrm{sd}$ is the strength of the local sd-coupling between the itinerant electrons and the localized magnetic moments. The vector $\vec S_i$ rotates with the period $\lambda$, i.e., 
\begin{align}
    \vec S_i =
    \begin{pmatrix}
    0\\
        \cos \left(\frac{2 \pi}{\lambda} x_i \right)
        \\
        \sin \left(\frac{2 \pi}{\lambda} x_i \right)
    \end{pmatrix},
\end{align}
where $x_i$ is the $x$-coordinate of lattice site $i$. 
The sites are equidistant, and the distance between the adjacent sites is denoted by $a$.

We generalize the model to two dimensions (2D) by stacking $N_y$ copies of the chain described by Eq.~\eqref{sm1-model-h-def}. This corresponds to a ferromagnetic interchain order considered in Ref.~\cite{hodt_arxiv_24}. We apply periodic boundary conditions in the $y$-direction, and open boundary conditions in the $x$-direction. Since the system is translationally invariant in the $y$-direction, we perform the discrete Fourier transform
\begin{align}
    c_{i_x, i_y, \sigma} = \frac{1}{\sqrt N_y} \sum_{i_y} c_{i_x, k_y, \sigma} e^{i k_y i_y a},
\end{align}
where the sum is over all sites in the $y$-direction, $N_y$ is the number of sites in the $y$-direction, and $a$ is the lattice constant.

\begin{figure}[htb] 
    \centering
    \subfloat[]{
    \includegraphics[width=0.49\linewidth]{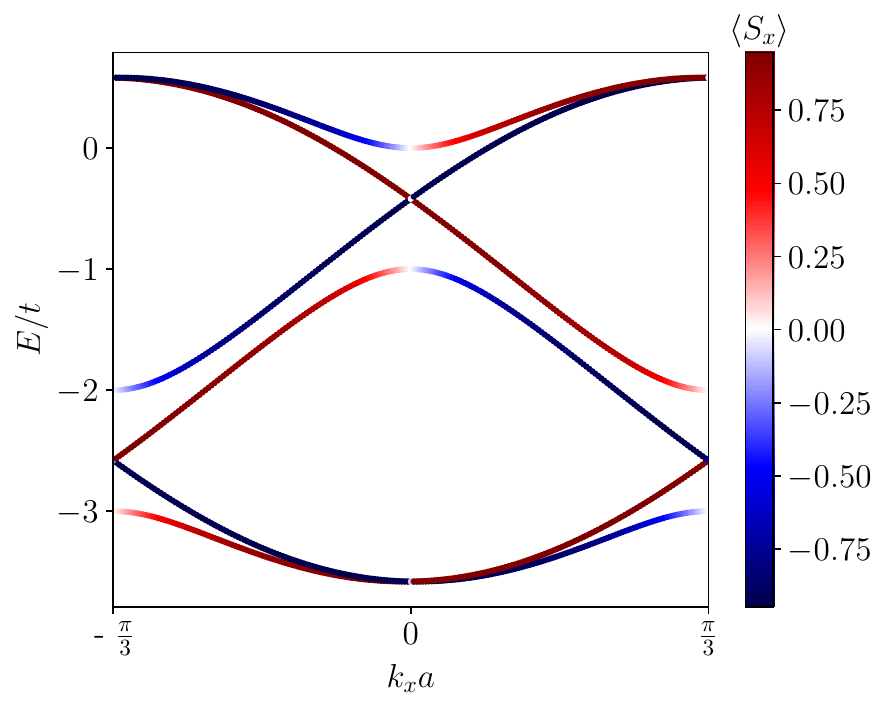}}
    \subfloat[]{
    \includegraphics[width=0.49\linewidth]{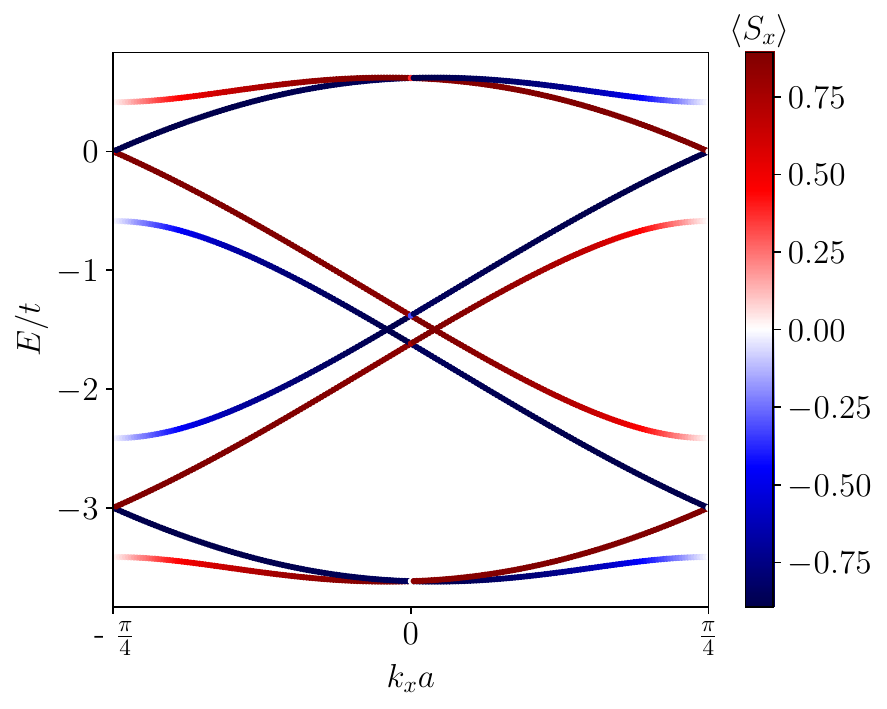}}
    \\
    \subfloat[]{
    \includegraphics[width=0.49\linewidth]{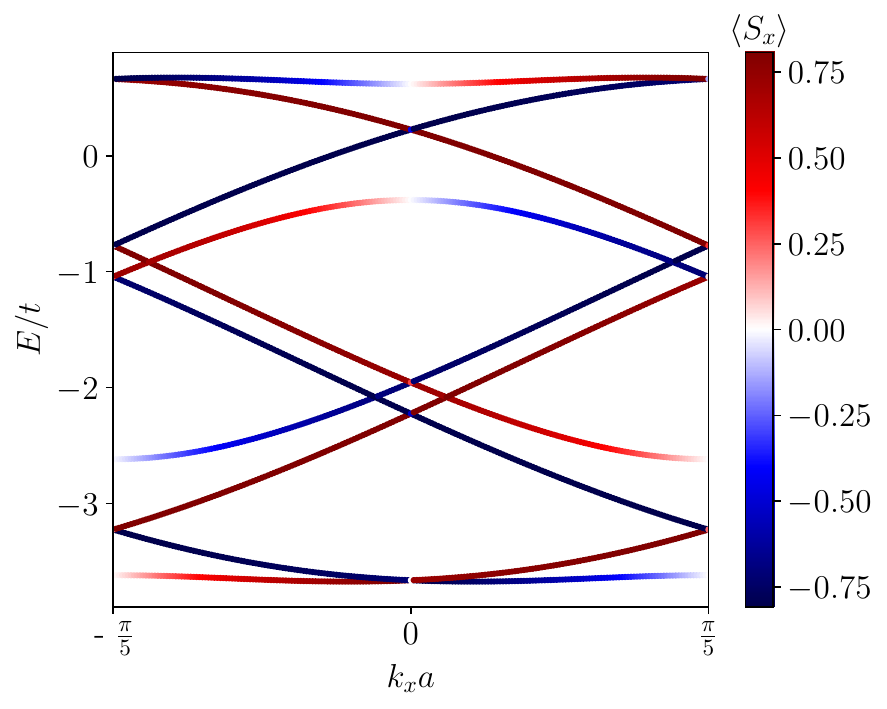}}
    \subfloat[]{
    \includegraphics[width=0.49\linewidth]{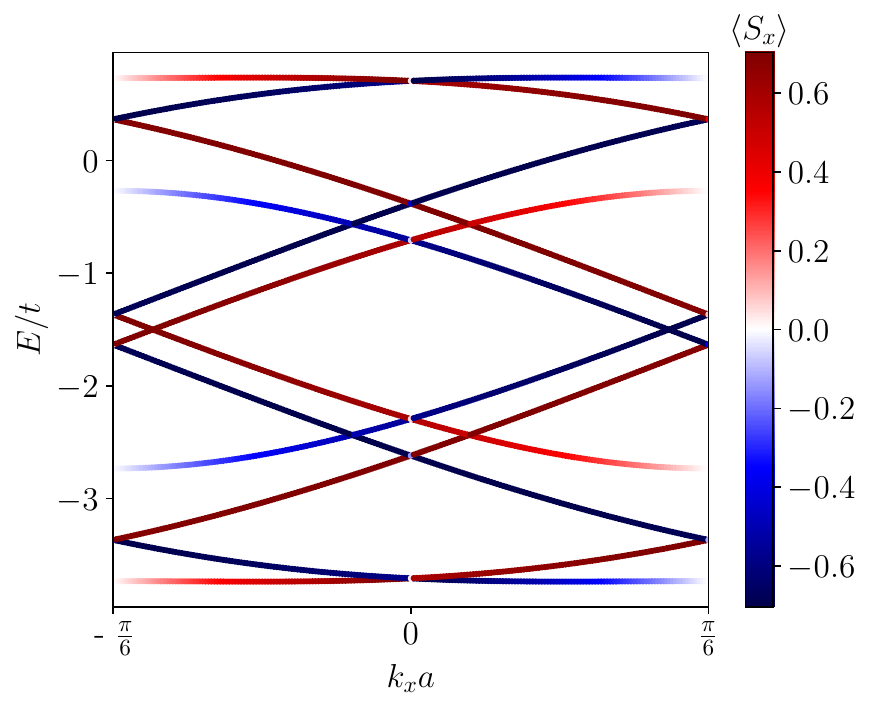}}
    \caption{The band structure for a bulk \pmag helimagnetic lattice model given in Eq.~\eqref{sm1-model-h-def}. The model parameters are $\mu = -2.5 t$, $J_\mathrm{sd} = 0.5 t$, and the period is $\lambda = \{3, 4, 5, 6\} a$ for subfigures (a)-(d), respectively.
    }
    \label{fig:bands}
\end{figure}

\begin{figure}[htb] 
    \centering
    \subfloat[]{
    \includegraphics[width=0.49\linewidth]{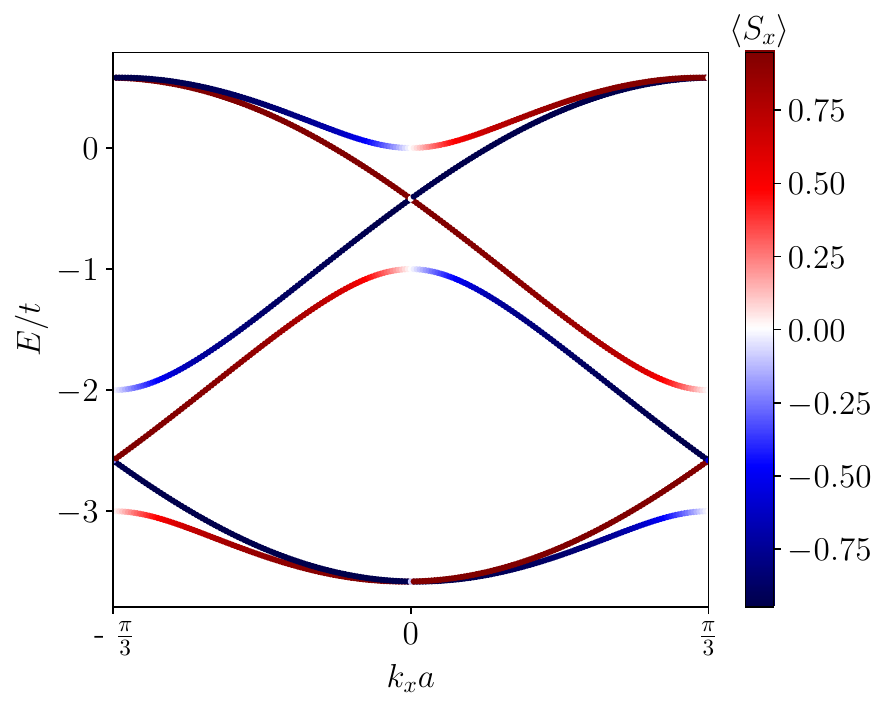}}
    \subfloat[]{
    \includegraphics[width=0.49\linewidth]{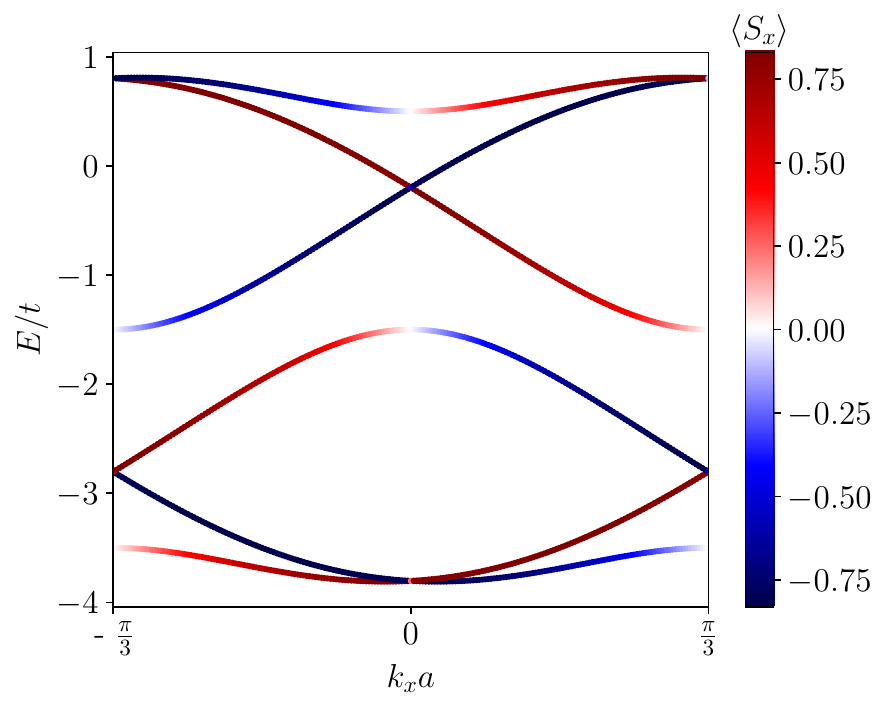}}
    \\
    \subfloat[]{
    \includegraphics[width=0.49\linewidth]{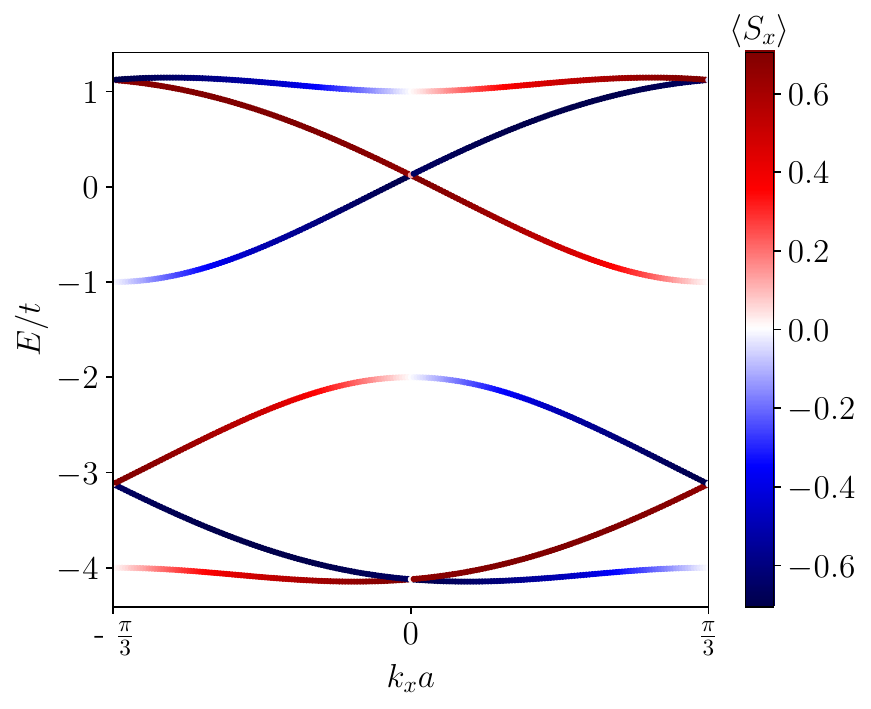}}
    \subfloat[]{
    \includegraphics[width=0.49\linewidth]{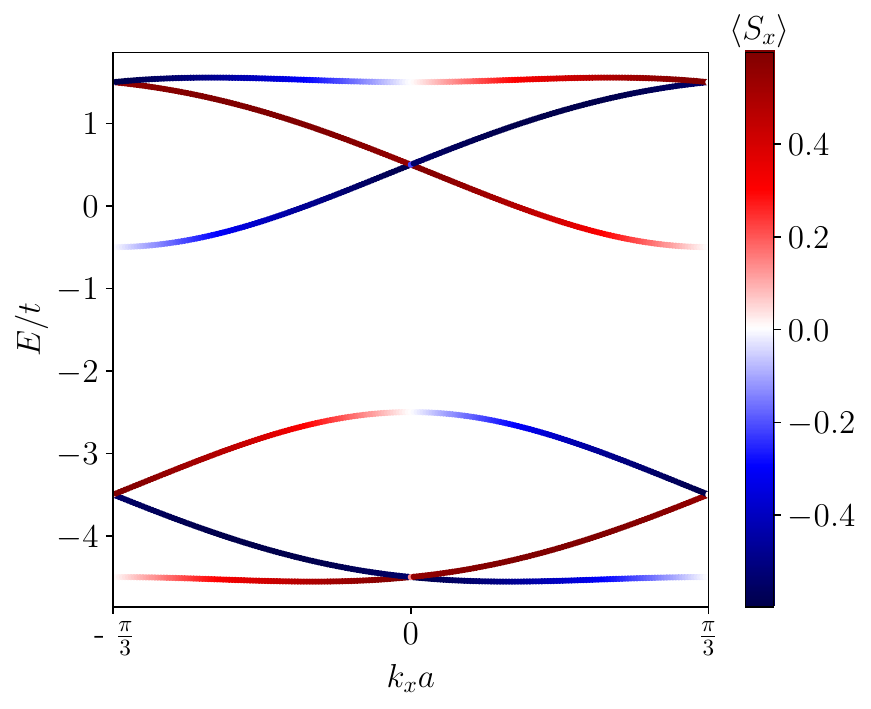}}
    \caption{
    The band structure for a bulk \pmag helimagnetic lattice model given in Eq.~\eqref{sm1-model-h-def}. The model parameters are $\mu = -0.5 t$, the period is $\lambda = 3 a$, and $J_\mathrm{sd} = \left\{0.5, 1.0, 1.5, 2.0\right\} t$ for subfigures (a)-(d), respectively.
    }
    \label{fig:bands-J}
\end{figure}

The rotating magnetization in the $yz$-plane results in a $p$-wave polarization in the $x$-direction. Considering a unit cell of the size of one helimagnetic period, we Fourier transform along the $x$-direction, resulting in a band structure with $2 \lambda/a$ bands. The band structure and polarization are shown for a few different values of $\lambda/a$ and $J_{\rm sd}/t$ in Figs. \ref{fig:bands} and \ref{fig:bands-J}, respectively. Note that since the size of the unit cell is $\lambda$, $k_x$ wavevector acquires smaller values, $k_x \in \big(-\pi / \lambda, \pi / \lambda \big]$.

We model the proximity-induced $p$-wave magnetism in a thin superconductor by adding a coexisting on-site BCS superconducting term
\begin{align}
    H_{\rm SC} =  - \sum_i (\Delta_i^{\vphantom{*}} c^\dag_{i\downarrow} c^\dag_{i\uparrow} 
    + \Delta_i^* c_{i\uparrow} c_{i\downarrow}),
\end{align}
where the order parameter is calculated self-consistently via the usual relation
\begin{align}
    \Delta_i = U \langle c_{i \uparrow} c_{i \downarrow} \rangle.
\end{align}
Here, $U > 0 $ is the strength of the on-site attractive potential, which is assumed to be constant throughout the material.
The critical temperature is found by a binomial search in temperatures, where, for each temperature, we check whether the median value of $\Delta_i$ increases or decreases after $5$ iterations. This is similar to Ref.~\cite{johnsenMagnetizationReorientationDue2019}. A similar procedure was also used to find the critical value of $J_\mathrm{sd}$ at zero temperature.

%% file: Supplemental-pMSC-2.tex
\subsection{Gap equation and free energy in the functional integral approach}
\label{sec:sm2-gap}

To address the intrinsic coexistence of $p$-wave magnetism with superconductivity, we solve the gap equation and select the solution that minimizes the free energy.

In the derivation of the gap equation, we use the functional integral approach, see, e.g., Refs.~\cite{Fossheim-Sudboe:book, Hugdal-Sudbo:2019}. We start with the following action in $d$-dimensional space:
\begin{eqnarray}
\label{2-model-S-def}
S &=& -\frac{1}{L^d T} \sum_{K} c_{K,\alpha}^{\dag} G_{0; \alpha \beta}^{-1}(K) c_{K,\beta} -\frac{1}{L^{3d} T^3} \sum_{K,K',Q} V_{K,K'; \alpha, \beta, \alpha', \beta'} c_{K'+Q/2,\alpha'}^{\dag} c_{-K'+Q/2,\beta'}^{\dag} c_{-K+Q/2,\beta}c_{K+Q/2,\alpha},
\end{eqnarray}
where $c_{K,\alpha}$ ($c_{K,\alpha}^{\dag}$) is the fermion annihilation (creation) operator with the spin $\alpha=\uparrow, \downarrow$, $K=\{i\omega_m, \mathbf{k}\}$, $\omega_m=(2m+1)\pi T$ are the Matsubara frequencies with $m \in\mathds{Z}$, $T$ is temperature, $\sum_{K} = \sum_{\omega_m}\sum_{\mathbf{k}}$, $L$ is the system size, and $G_{0}^{-1}(K)$ is the inverse Green's function of a noninteracting system. We assume that the superconducting pairing potential can be factorized as $V_{K,K'; \alpha, \beta, \alpha', \beta'} = g v_{\alpha \beta}(K)v_{\alpha' \beta'}(K')$~\cite{Fossheim-Sudboe:book} with $g$ being the interaction strength. 
The momentum structure of the pairing interaction is quantified by the matrices $v(K)$ and $v(K')$; we suppress the indices when this does not lead to confusion. 

The full partition function is
\begin{equation}
\label{2-model-Z}
\mathcal{Z} = \int Dc^{\dag}Dc\, e^{-S}.
\end{equation}
By performing the Hubbard-Stratonovich transformation, we rewrite the four-fermion term, i.e., the last term in Eq.~\eqref{2-model-S-def}, in terms of the bosonic fields $\phi_Q$:
\begin{eqnarray}
\label{2-model-HS-1}
&&e^{\frac{1}{L^{3d} T^3} \sum_{K,K',Q} g v_{\alpha \beta}(K)v_{\alpha' \beta'}(K') c_{K'+Q/2,\alpha'}^{\dag} c_{-K'+Q/2,\beta'}^{\dag} c_{-K+Q/2,\beta}c_{K+Q/2,\alpha}} \nonumber\\
&&=\int D\phi_{Q}^{\dag}D\phi_{Q} e^{-\frac{L^d}{g T} \sum_{Q} |\phi_{Q}|^2 -\frac{T}{L^d}\sum_{K',Q} v_{\alpha' \beta'}(K') \left[\phi_{Q} c^{\dag}_{K'+Q/2,\alpha'} c_{-K'+Q/2,\beta'}^{\dag} +h.c.\right]}.
\end{eqnarray}
Therefore, the corresponding effective action is
\begin{equation}
\label{2-model-S-eff-1}
S_{\rm eff} = -\frac{T}{L^d} \sum_{K} c^{\dag}_{K} G_{0}^{-1}c_{K}
-\frac{T}{L^d}\sum_{K,Q} v_{\alpha \beta}(K) \left[\phi_{Q} c^{\dag}_{K+Q/2,\alpha} c_{-K+Q/2,\beta}^{\dag} +h.c.\right]
+\sum_{Q} \frac{L^d}{g T} |\phi_{Q}|^2.
\end{equation}

To describe superconductivity, it is convenient to work in the Nambu space. We define the corresponding bi-spinor as
\begin{equation}
\label{2-model-Nambu}
\Psi_{K} = \left\{c_{K,\uparrow}, c_{K,\downarrow}, c_{-K,\uparrow}^{\dag}, c_{-K, \downarrow}^{\dag}\right\}^{T}.
\end{equation}
Then, the effective action \eqref{2-model-S-eff-1} can be rewritten in a compact form
\begin{equation}
\label{2-model-S-eff-2}
S_{\rm eff} = -\frac{T}{2L^d} \sum_{K,K'} \Psi^{\dag}_{K} \hat{G}_{K,K'}^{-1} \Psi_{K'} +\frac{L^d}{g T} \sum_{Q} |\phi_{Q}|^2,
\end{equation}
where
\begin{equation}
\label{2-model-G-Nambu}
\hat{G}^{-1}(K,K') =
\begin{pmatrix}
G_{0}^{-1}(K+Q/2) \delta_{K,K'} & v(K) \phi_{K-K'} \\
v^T(K) \phi^{*}_{K'-K} & -\delta_{K,K'} \left[G_{0}^{-1}(-K+Q/2)\right]^{T}
\end{pmatrix}
\end{equation}
is the inverse full Green function.

By integrating out fermions in the effective action \eqref{2-model-S-eff-2}, we derive the following action for the bosonic field $\phi_Q$~\cite{Altland:2010-book}:
\begin{equation}
\label{2-model-S-phi}
S_{\phi} = \frac{L^d}{g T} \sum_{Q} |\phi_{Q}|^2 - \frac{1}{2} \tr{\mbox{ln}{\left(G^{-1}\right)}}.
\end{equation}

We focus on the uniform pairing and assume a spatially homogeneous field $\phi_{Q}=\delta_{Q,0}\phi$. The free energy of the system is defined as
\begin{equation}
\label{2-model-F}
\frac{F}{T} = \frac{L^d}{T} \frac{|\phi|^2}{g} -\frac{1}{2} \tr{\,\mbox{ln}{\left(G^{-1}\right)}}
=\frac{L^d}{T} \frac{|\phi|^2}{g} -\frac{1}{2} \mbox{ln}{\,\mbox{det} G^{-1}}
=\frac{L^d}{T} \frac{|\phi|^2}{g} -\frac{1}{2} \sum_{K} \sum_j \ln{\left[\lambda_j(K)\right]},
\end{equation}
where $\lambda_j(K)$ are eigenvalues of $G^{-1}(K)$. In the case of energy-independent interactions $\lambda_j(K)=i\omega_m +\lambda(\mathbf{k})$. Then, the summation over the Matsubara frequencies can be performed: $T\sum_{\omega_m} \ln{\left(-i\omega_m +\epsilon\right)} = T \ln{\left(1 + e^{-\epsilon/T}\right)}$. This allows to simply Eq.~(\ref{2-model-F}) as
\begin{equation}
\label{2-model-F-1}
\frac{F}{T} =\frac{L^d}{T} \frac{|\phi|^2}{g} -\frac{1}{2} \sum_{\mathbf{k}} \sum_j \ln{\left(1+e^{\lambda_j(\mathbf{k})/T}\right)} +\mbox{const}.
\end{equation}
To determine whether the superconducting state is energetically favorable, we subtract the normal-state free energy, which is obtained by setting $\phi=0$,
\begin{equation}
\label{2-model-F-S-N}
\delta F = \frac{F_S-F_N}{L^d} =\frac{|\phi|^2}{g} -\frac{T}{2} \frac{1}{L^d} \sum_{\mathbf{k}} \sum_j \ln{\left(\frac{1+e^{\lambda_j(\mathbf{k})/T}}{1+e^{\lambda^{(N)}_j(\mathbf{k})/T}}\right)},
\end{equation}
where $\lambda^{(N)}_j(\mathbf{k}) = \lambda_j(\mathbf{k})\big|_{\phi\to0}$. 

Variating the free energy difference in Eq.~\eqref{2-model-F-S-N} with respect to $\phi^{*}$, we obtain the following gap equation:
\begin{equation}
\label{2-model-gap-phiQ}
\phi = \frac{g}{2} \frac{1}{L^d} \sum_{\mathbf{k}} \sum_{j} \frac{\partial \lambda_j(\mathbf{k})}{\partial \phi^{*}} \frac{1}{1+e^{-\lambda_j(\mathbf{k})/T}}.
\end{equation}

\subsection{Effective model of \pmag}
\label{sec:sm2-model}

Before discussing the solutions of the gap equation, let us show how the effective low-energy model used in the main text can be derived from the \pmag model advocated in Ref.~\cite{brekke_prl_24}, see the SM there. In the vicinity of the $\Gamma$-point, we obtain the following Hamiltonian:
\begin{equation}
\label{2-model-full}
H_{\rm eff}(\mathbf{k}) = \xi_k\tau_0 \otimes \sigma_0
    + \frac{\left(\bm{\alpha}\cdot \mathbf{k}\right)}{2m} \tau_0 \otimes \sigma_{z'}  + J \tau_z \otimes \sigma_{x'}  + t_{\rm inter} \tau_x \otimes \sigma_0,
\end{equation}
where $\xi_k=k^2/(2m) -\mu$, $m$ is the effective mass, $\mu$ is the Fermi energy, $\bm{\alpha}$ is the spin-spliting vector, $J$ is the sd coupling strength, $t_{\rm inter}$ is the inter-sectoral~\footnote{Note that while the introduction of the sectoral degree of freedom provides an intuitive illustration of the physical meaning of the model, it is not strictly necessary, and is flexible to accommodate any low-energy bands in $p$-wave magnets regardless of their origin.} coupling, and we interchanged sectoral $\bm{\tau}$ and spin $\bm{\sigma}$ matrices compared to Ref.~\cite{brekke_prl_24}. The sectors of the effective model host localized spins that are related by the combined time reversal and translation $\mathcal{T}\bm{\tau}$ symmetry. The spin-dependent intra-sectoral hopping quantified by $\bm{\alpha}$ is a necessary ingredient for the realization of the $p$-wave spin polarization. While not being crucial for the $p$-wave spectrum, the inter-sectoral hopping is also present and can be physically motivated by the hopping between the nearest neighbors in the lattice model. The spin coordinates are primed to show that they are decoupled from the crystal coordinates. Therefore, the effective model \eqref{2-model-full} is universal and is able to describe different spin textures. We would also like to note that, constructed via symmetry considerations rather than directly from the tight-binding Hamiltonian, the effective continuum model contains more parameters than its lattice counterpart. This makes the effective model more flexible in describing different types of $p$-wave spectra.

\begin{figure}[ht!]
\centering
\begin{subfloat}[]{\includegraphics[height=0.21\columnwidth]{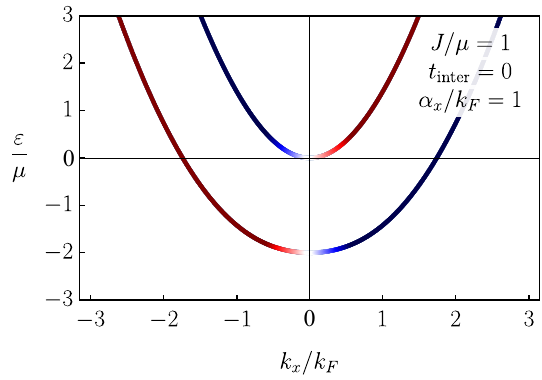}}
\end{subfloat}
\begin{subfloat}[]{\includegraphics[height=0.21\columnwidth]{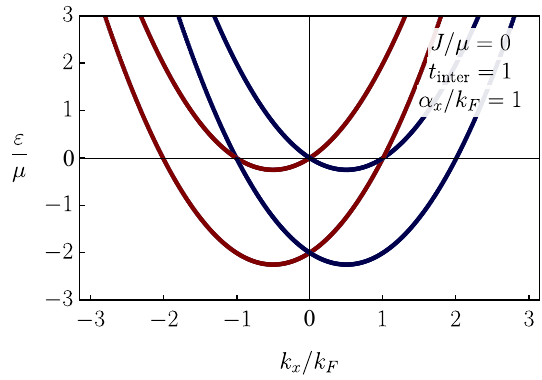}}
\end{subfloat}
\begin{subfloat}[]{\includegraphics[height=0.21\columnwidth]{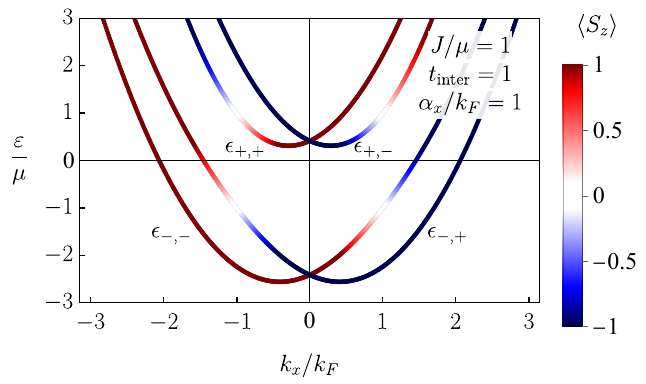}}
\end{subfloat}
\caption{\label{fig:2-model-full-eps-bands}
The energy spectrum of the full model given in \eqref{2-model-full-eps} where the color scale corresponds to the spin polarizarion $\left\langle S_z \right\rangle = \Psi_{s_1,s_2}^{\dag}\hat{s}_z\Psi_{s_1,s_2}$ (in the units of $\hbar/2$). In all panels, we fix $\alpha_x=k_F$, $\alpha_y=0$, and $k_F=\sqrt{2m \mu}$.
}
\end{figure}

The energy spectrum of the model \eqref{2-model-full} contains four bands
\begin{eqnarray}
\label{2-model-full-eps}
\epsilon_{k, s_1, s_2} = \xi_k +s_1 \sqrt{J^2 +\left[\left(\bm{\alpha}\cdot \mathbf{k}\right)/(2m) +s_2 t_{\rm inter}\right]^2},
\end{eqnarray}
where $s_1=\pm$ and $s_2=\pm$.
We find it convenient to number the states as $\epsilon_{1}=\epsilon_{k, -, +}$, $\epsilon_{2}=\epsilon_{k, -, -}$, $\epsilon_{3}=\epsilon_{k, +, -}$, and $\epsilon_{4}=\epsilon_{k, +, +}$. We show the spectrum \eqref{2-model-full-eps} in Fig.~\ref{fig:2-model-full-eps-bands}. In the presence of the inter-sectoral coupling, the lower two bands resemble those for a Rashba model albeit with the spin polarization along the $z$-direction, see Fig.~\ref{fig:2-model-full-eps-bands}(b). The intra-sectoral hopping makes the spin texture nontrivial, see Figs.~\ref{fig:2-model-full-eps-bands}(a) and \ref{fig:2-model-full-eps-bands}(c).

In the case of well-separated low- and high-energy bands, it is possible to simplify the model \eqref{2-model-full} further. By projecting out high-energy bands, the low-energy Hamiltonian is defined as
\begin{equation}
\label{2-model-h-le-def}
H_{\rm le}(\mathbf{k}) =
\begin{pmatrix}
\Psi^{\dag}_{-,+} H_{\rm eff}(\mathbf{k}) \Psi_{-,+} & \Psi^{\dag}_{-,+} H_{\rm eff}(\mathbf{k}) \Psi_{-,-} \\
\Psi^{\dag}_{-,-} H_{\rm eff}(\mathbf{k}) \Psi_{-,+} & \Psi^{\dag}_{-,-} H_{\rm eff}(\mathbf{k}) \Psi_{-,-}
\end{pmatrix},
\end{equation}
where we used the normalized wave function
\begin{eqnarray}
\label{2-model-Psi-def}
\Psi_{s_1,s_2} &=& \frac{|J|}{2 \sqrt{\left[(\bm{\alpha}\cdot\mathbf{k})/(2m) +t_{\rm inter}\right] \left[(\bm{\alpha}\cdot\mathbf{k})/(2m) +t_{\rm inter} +\epsilon_{s_1,s_2}-\xi_k\right] +J^2}}
\begin{pmatrix}
-\frac{(\bm{\alpha}\cdot\mathbf{k})/(2m) +t_{\rm inter} +\epsilon_{s_1,s_2}-\xi_k}{J}\\
-1\\
-\frac{(\bm{\alpha}\cdot\mathbf{k})/(2m) +t_{\rm inter} +\epsilon_{s_1,s_2}-\xi_k}{J}\\
1
\end{pmatrix}.
\end{eqnarray}

Assuming the strong inter-sectoral coupling $\sqrt{t_{\rm inter}^2+J^2} \gg \alpha k_F$, Eq.~\eqref{2-model-h-le-def} is simplified as
\begin{eqnarray}
\label{2-model-h-le}
H_{\rm le}(\mathbf{k}) &=&
\begin{pmatrix}
\xi_k -\sqrt{J^2 +\left[t_{\rm inter} -\left(\bm{\alpha}\cdot \mathbf{k}\right)/(2m)\right]^2} & 0 \\
0 & \xi_k -\sqrt{J^2 +\left[t_{\rm inter} +\left(\bm{\alpha}\cdot \mathbf{k}\right)/(2m)\right]^2}
\end{pmatrix} \nonumber\\
&\approx& \xi_k -\tilde{J} +\frac{\left(\bm{\zeta}\cdot \mathbf{k}\right)}{m} \rho_z
= \frac{1}{2m}\left(\mathbf{k} +\rho_z\bm{\zeta}\right)^2 -\mu -\tilde{J} -\frac{\zeta^2}{2m},
\end{eqnarray}
where $\rho_z$ is the Pauli matrix in the band space. To simplify the notations, we introduce $\tilde{J}=\sqrt{J^2 +t_{\rm inter}^2}$ and $\bm{\zeta} = t_{\rm inter} \bm{\alpha}/(2\tilde{J})$.

The eigenvalues and eigenfunctions of the model \eqref{2-model-h-le} are
\begin{eqnarray}
\label{2-model-eps-le}
\epsilon_{s} &=& \xi_k -\tilde{J} +s\frac{\left(\bm{\zeta}\cdot \mathbf{k}\right)}{m}
= \frac{1}{2m}\left(\mathbf{k} +s \bm{\zeta}\right)^2 -\mu -\tilde{J} -\frac{\zeta^2}{2m},\\
\label{2-model-psi-le}
\psi_{s} &=& \left\{\frac{1 +s}{2}, \frac{1 -s}{2}\right\}^{T},
\end{eqnarray}
where $s=\pm$.
As can be seen from Eq.~\eqref{2-model-eps-le}, the energy spectrum is composed of two shifted parabolas with minima at $\mathbf{k} = -s\bm{\zeta}$. The eigenfunctions are trivial in the band space and are the same as for any two-band model with full spin polarization.

The normalized by $\hbar/2$ spin operator $\hat{s}_z=\tau_0\otimes\sigma_z$ in the basis of the Hamiltonian \eqref{2-model-h-le} reads as
\begin{equation}
\label{2-model-sigmaz-le}
\hat{s}_z =
\begin{pmatrix}
-\frac{\left(\bm{\alpha}\cdot \mathbf{k}\right)/(2m) -t_{\rm inter}}{\sqrt{J^2 +\left[\left(\bm{\alpha}\cdot \mathbf{k}\right)/(2m) -t_{\rm inter}\right]^2}} & 0 \\
0 & -\frac{\left(\bm{\alpha}\cdot \mathbf{k}\right)/(2m) +t_{\rm inter}}{\sqrt{J^2 +\left[\left(\bm{\alpha}\cdot \mathbf{k}\right)/(2m) +t_{\rm inter}\right]^2}}
\end{pmatrix}
\approx -\frac{\left(\bm{\zeta}\cdot \mathbf{k}\right) J^2}{m \tilde{J}^{2}} \rho_0 + \frac{t_{\rm inter}}{\tilde{J}} \rho_z \approx \frac{t_{\rm inter}}{\tilde{J}} \rho_z,
\end{equation}
where we expanded up to the leading nontrivial order in $\left(\bm{\zeta}\cdot \mathbf{k}\right)$ in the second expression and kept only the lowest-order term in the last expression. Therefore, up to a prefactor, the spin and band degrees in the low-energy effective model \eqref{2-model-h-le} coincide in the low-energy model.

As we mentioned in the main text, the Hamiltonian \eqref{2-model-h-le} is qualitatively similar to that used in Refs.~\cite{maedaTheoryTunnelingSpectroscopy2024, salehi_arxiv_24, Ezawa-PurelyElectricalDetection-2024, fukaya_arxiv_24, ezawaThirdorderFifthorderNonlinear2024}. The latter resembles the electron Hamiltonian for the persistent spin helix~\cite{Schliemann-Loss-NonballisticSpinFieldEffectTransistor-2003, Bernevig-Zhang-ExactSUSymmetry-2006, Schliemann-Schliemann-ColloquiumPersistentSpin-2017}. The only technical difference is in the quadratic in the spin-splitting $\zeta$ term in Eq.~\eqref{2-model-h-le}. However, this term acts as an effective chemical potential, and, therefore, does not play crucial role.

\subsection{Pairing interaction in the diagonal basis}
\label{sec:sm2-pairing-diag}

Intrinsic superconductivity relies on the pairing of long-lived quasiparticles, i.e., the pairing in the diagonal basis. Let us show that such a pairing for the simplest phonon-mediated interaction also acquires a simple form. 

The interacting Hamiltonian for the full effective model \eqref{2-model-full} is
\begin{equation}
\label{2-model-full-Hint}
H_{\text{int}} = -\frac{V}{N}\sum_{kk'} c_{k,\sigma,\eta}^\dagger c_{-k,-\sigma,\eta'}^\dagger c_{-k',-\sigma,\eta} c_{k',\sigma,\eta'},
\end{equation}
where we assumed spin-singlet pairing and $\eta$ is the sectoral degree of freedom. 

The original basis is related to the basis that diagonalizes the effective Hamiltonian as $c_{k} = \hat{M} \gamma_k$, where the matrix $\hat{M}=\left\{\Psi_{k,-,+}, \Psi_{k,-,-}, \Psi_{k,+,-}, \Psi_{k,+,+}\right\}$ is composed of the eigenvectors given in Eq.~\eqref{2-model-Psi-def}. By using this relation in \eqref{2-model-full-Hint}, the interacting Hamiltonian in the diagonal basis can be derived. The corresponding expression is, however, bulky and not too informative. In what follows, we analyze a few limiting cases.

Making the same assumptions as in the derivation of the Hamiltonian \eqref{2-model-h-le}, i.e., assuming that strong inter-sectoral coupling $t_{\rm inter} \gg J, \alpha k_F$, we obtain
\begin{equation}
\label{2-model-full-Hint-diag-tinter}
H_{\text{int}} \approx  -\frac{V}{4N}\sum_{kk'}
\left(\gamma_{2,k} -\gamma_{4,k}\right)^{\dag} \left(\gamma_{1,-k} +\gamma_{3,-k}\right)^{\dag} \left(\gamma_{2,-k'} -\gamma_{4,-k'}\right) \left(\gamma_{1,k'} +\gamma_{3,k'}\right) \approx -\frac{V}{4N}\sum_{kk'}
\gamma_{2,k}^{\dag} \gamma_{1,-k}^{\dag} \gamma_{2,-k'} \gamma_{1,k'}.
\end{equation}
Formally, there are interactions between the bands $1$ and $4$. These bands, however, are well-separated for $t_{\rm inter}\gg J, \alpha k_F$ with the bands $3$ and $4$ being above the Fermi energy, see, e.g., Fig.~\ref{fig:2-model-full-eps-bands}(c). Omitting these bands, we obtain a simple interaction Hamiltonian of the same form as for the spin-singlet pairing in the original basis, see the last expression in Eq.~\eqref{2-model-full-Hint-diag-tinter}. Note that while the leading-order in $J/t_{\rm inter}, \alpha k_F/t_{\rm inter} \ll1$ term is independent of whether we pair between the same or different sectors, a discrepancy appears in the first order in $J/t_{\rm inter}, \alpha k_F/t_{\rm inter}  \ll1$.

Expanding in $J\gg t_{\rm inter}, \alpha k_F$ and neglecting the upper ($3$ and $4$ bands), we obtain
\begin{equation}
\label{2-model-full-Hint-diag-J-intra}
H_{\text{int}} \approx -\frac{V}{16N}\sum_{kk'}
\left(\gamma_{1,k} -\gamma_{2,k}\right)^{\dag} \left(\gamma_{1,-k} -\gamma_{2,-k}\right)^{\dag} \left(\gamma_{1,-k'} -\gamma_{2,-k'}\right) \left(\gamma_{1,k'} -\gamma_{2,k'}\right)
\end{equation}
for intra-sectoral pairing and
\begin{equation}
\label{2-model-full-Hint-diag-J-inter}
H_{\text{int}} \approx -\frac{V}{16N}\sum_{kk'}
\left(\gamma_{1,k} -\gamma_{2,k}\right)^{\dag} \left(\gamma_{1,-k} +\gamma_{2,-k}\right)^{\dag} \left(\gamma_{1,-k'} -\gamma_{2,-k'}\right) \left(\gamma_{1,k'} +\gamma_{2,k'}\right)
\end{equation}
for the inter-sectoral pairing. There are also terms corresponding to the inter-band pairing.

Thus, the electron-phonon interaction results in the inter-band interaction for the full effective model \eqref{2-model-full} and the low-energy effective model \eqref{2-model-h-le}.

\subsection{Solutions to the gap equations}
\label{sm2-sol}

To solve the gap equation \eqref{2-model-gap-phiQ} and calculate the free energy difference \eqref{2-model-F-S-N}, we need to determine the eigenvalues of the Green's functions $\lambda_j(\mathbf{k})$. In the low-energy model \eqref{2-model-h-le}, the Green's function $G_0(K)$ is defined as
\begin{equation}
\label{2-model-G0-def}
G_{0}(i\omega_m, \mathbf{k}) = \left[i\omega_m -H_{\rm le}(\mathbf{k})\right]^{-1}.
\end{equation}
Then, for uniform pairing, the full Green's function \eqref{2-model-G-Nambu} reads
\begin{equation}
\label{2-model-G-Nambu-MF}
\hat{G}^{-1}(K,K') = \delta_{K,K'}
\begin{pmatrix}
G_{0}^{-1}(K)  & \hat{\Delta}(\mathbf{k}) \\
\hat{\Delta}^{\dag}(\mathbf{k}) & - \left[G_{0}^{-1}(-K)\right]^{T}
\end{pmatrix},
\end{equation}
where $\hat{\Delta}(\mathbf{k}) = v(\mathbf{k}) \left[\Delta_0 +(\bm{\rho}\cdot \bm{\Delta})\right] (i\rho_y)$. Here, $\Delta_0$ corresponds to the spin-singlet gap, $\Delta_z$ is the mixed-spin spin-triplet gap, and $\Delta_{x,y}$ are equal-spin spin-triplet gaps. For spatially uniform gaps, the equal-spin triplet gaps should vanish. Note, also, that the band and spin bases are equal for the low-energy effective model \eqref{2-model-h-le}.

The eigenvalues for the gaps $\Delta_0$ and $\Delta_z$ are
\begin{equation}
\label{2-model-lambda-03}
\lambda_{s_1,s_2} = s_1 \sqrt{\left(\frac{k_{s_2}^2}{2m} -\mu -\tilde{J} -\frac{\zeta^2}{2m} \right)^2 +|v(\mathbf{k})|^2 |\Delta_{0,z}|^2},
\end{equation}
where we introduced $\mathbf{k}_{s} = \mathbf{k} +s \bm{\zeta}$.

\subsubsection{$s$-wave spin-singlet}
\label{sec:2-model-uniform-s}

In the case of $s$-wave gaps, $v(\mathbf{k})=1$. The spin-splitting parameter $\bm{\zeta}$ can be absorbed into the spin-dependent $\xi_{k,s}$ leaving only a term $\propto \zeta^2$ that modifies the chemical potential. Therefore, as long as this modification is small, we do not expect any drastic difference from a BCS superconductor. Indeed, in the limit $T\to0$, the gap equation \eqref{2-model-gap-phiQ} reads
\begin{eqnarray}
\label{2-model-gap-eq-regular}
1 &=& \frac{g}{2} \sum_{s_1,s_2=\pm}\int \frac{d\mathbf{k}}{(2\pi)^2} \frac{\Theta{\left(\lambda_{s_1,s_2}\right)}}{2\lambda_{s_1,s_2}}
=\frac{g}{4} \sum_{s_2=\pm}\int \frac{d\mathbf{k}_{s_2}}{(2\pi)^2} \frac{1}{\sqrt{\left(\frac{k_{s_2}^2}{2m} -\mu -\tilde{J} -\frac{\zeta^2}{2m} \right)^2 +|\Delta_{0}|^2}} \nonumber\\
&=& g \frac{\nu_0}{4} \sum_{s_2=\pm} \int_0^{\infty} d\left(\frac{k_{s_2}^2}{2m}\right) \int_0^{2\pi} \frac{d\theta}{2\pi} \frac{1}{\sqrt{\left(\frac{k_{s_2}^2}{2m} -\mu -\tilde{J} -\frac{\zeta^2}{2m} \right)^2 +|\Delta_{0}|^2}} \nonumber\\
&=&g \frac{\nu_0}{4} \sum_{s_2=\pm} \int_{-\mu -\tilde{J} -\frac{\zeta^2}{2m} }^{\infty} d\left(\frac{k_{s_2}^2}{2m} -\mu -\tilde{J} -\frac{\zeta^2}{2m}  \right)  \int_0^{2\pi} \frac{d\theta}{2\pi} \frac{1}{\sqrt{\left(\frac{k_{s_2}^2}{2m} -\mu -\tilde{J} -\frac{\zeta^2}{2m} \right)^2 +|\Delta_{0}|^2}} \nonumber\\
&\approx& g \frac{\nu_0}{4} \sum_{s_2=\pm} \int_{-\infty}^{\infty} d \xi_{k,s_2} \int_0^{2\pi} \frac{d\theta}{2\pi} \frac{1}{\sqrt{\xi_{k,s_2}^2 +|\Delta_{0}|^2}}
\approx g \frac{\nu_0}{4} \sum_{s_2=\pm} \int_{-\omega_D}^{\omega_D} d \xi_{k,s_2} \int_0^{2\pi} \frac{d\theta}{2\pi} \frac{1}{\sqrt{\xi_{k,s_2}^2 +|\Delta_{0}|^2}} \nonumber\\
&=& \frac{g\nu_0}{2} \ln{\left(\frac{\sqrt{\omega_D^2 +|\Delta_0|^2} +\omega_D}{\sqrt{\omega_D^2 +|\Delta_0|^2} -\omega_D}\right)}
\approx \tilde{g} \ln{\left(\frac{2\omega_D}{|\Delta_0|}\right)}
\end{eqnarray}
leading to the following standard solution for the gap:
\begin{equation}
\label{2-model-gap-regular}
|\Delta_0| = 2\omega_D e^{-1/(g\nu_0)}.
\end{equation}
We redefined momentum $\mathbf{k} \to \mathbf{k}_{s} = \mathbf{k} +s \bm{\zeta}$ in the first line in Eq.~\eqref{2-model-gap-eq-regular}, assumed that $\mu +\tilde{J} +\frac{\zeta^2}{2m} \gg |\Delta_0|$ in the fourth line, as well as introduced the energy cutoff $\omega_D$ in the same line. The cutoff $\omega_D$ limits interactions to the vicinity of the Fermi surface of the normal-state long-lived quasiparticles. In addition, $\nu_0=m/(2\pi)$ is the normal-state density of states (DOS).

The expression for the critical temperature also acquires a similar to the standard BCS result form:
\begin{eqnarray}
\label{2-model-gap-eq-regular-Tcr}
1 &=& \frac{g}{2} \sum_{s_1,s_2=\pm}\int \frac{d\mathbf{k}}{(2\pi)^2} \frac{1}{2\lambda_{s_1,s_2}} \frac{1}{1 +e^{-\lambda_{s_1,s_2}/T}} \nonumber\\
&=& g\frac{\nu_0}{2} \sum_{s_1=\pm} \sum_{s_2=\pm} \int_0^{\infty} d\left(\frac{k_{s_2}^2}{2m}\right) \int_0^{2\pi} \frac{d\theta}{2\pi} \frac{v^2(\theta)}{2s_1 \sqrt{\left(\frac{k_{s_2}^2}{2m} -\mu -\tilde{J} -\frac{\zeta^2}{2m} \right)^2 +v^2(\theta)|\Delta|^2}} \nonumber\\
&\times& \frac{1}{1 +e^{-s_1 \sqrt{\left(\frac{k_{s_2}^2}{2m} -\mu -\tilde{J} -\frac{\zeta^2}{2m} \right)^2 +v^2(\theta)|\Delta|^2}/T}} \nonumber\\
&\stackrel{T\to T_{\rm cr}}{\approx}& g\frac{\nu_0}{2}  \sum_{s_1=\pm} \sum_{s_2=\pm} \int_{-\infty}^{\infty} d\xi_{k,s_2} \int_0^{2\pi} \frac{d\theta}{2\pi} \frac{v^2(\theta)}{2s_1 |\xi_{k,s_2}|} \frac{1}{1 +e^{-s_1 |\xi_{k,s_2}|/T_{\rm cr}}}\nonumber\\
&=& g\frac{\nu_0}{2}   \sum_{s_2=\pm} \int_{-\infty}^{\infty} d\xi_{k,s_2} \int_0^{2\pi} \frac{d\theta}{2\pi} \frac{v^2(\theta)}{2 |\xi_{k,s_2}|} \tanh{\left(\frac{|\xi_{k,s_2}|}{2T_{\rm cr}}\right)} = g\nu_0 \int_{0}^{\omega_D}\! d\xi \frac{1}{|\xi|} \tanh{\left(\frac{|\xi|}{2T_{\rm cr}}\right)}
\nonumber\\
&=& g\nu_0  \left\{ \tanh{\left(\frac{|\xi|}{2T_{\rm cr}}\right)} \ln{\left(\frac{|\xi|}{T_{\rm cr}}\right)} \Big|_{0}^{\omega_{D}}\!
-\int_0^{\omega_{D}} \frac{d\xi}{2T_{\rm cr}} \frac{\ln{\left(\frac{|\xi|}{T_{\rm cr}}\right)}}{\cosh^2{[|\xi|/(2T_{\rm cr})]}}
\right\} \nonumber\\
&\approx& g\nu_0  \left\{ \ln{\left(\frac{\omega_{D}}{T_{\rm cr}}\right)}
-\int_0^{\infty} \frac{d\tilde{\xi}}{2} \frac{\ln{\tilde{\xi}}}{\cosh^2{\left(\tilde{\xi}/2\right)}}
\right\} = \ln{\left(\frac{\omega_D}{T_{\rm cr}} \frac{2 e^{\gamma_e}}{\pi}\right)},
\end{eqnarray}
where we used $\int_{0}^{2\pi}d\theta\, v^2(\theta)/(2\pi) = 1$ in the fifth line. In the fourth line, we used the fact that $\Delta\to0$ at $T \to T_{\rm cr}$. In the last expression, we assumed that $\omega_{D}/T_{\rm cr} \gg 1$ and used $\gamma_e\approx 2.72$ as the Euler constant. Using Eq.~\eqref{2-model-gap-eq-regular-Tcr}, it is straightforward to derive the standard BCS result for the critical temperature:
\begin{equation}
\label{2-model-gap-eq-regular-Tcr-fin}
T_{\rm cr} = \frac{2\omega_{D}}{\pi} e^{\gamma_e} e^{-1/(g \nu_0)}.
\end{equation}
 
A similar line of thought can be applied to the free energy difference given in  Eq.~\eqref{2-model-F-S-N}, which at $T\to 0$ reads
\begin{eqnarray}
\label{2-model-F-S-N-2}
\delta F &=&\frac{|\Delta_0|^2}{g} - \sum_{s_2}\int_0^{\infty} \frac{k_{s_2} dk_{s_2}}{4\pi} \int_0^{2\pi} \frac{d\theta}{2\pi}
\left(\lambda_{+,s_2} -\lambda_{+,s_2}^{(N)}\right) \nonumber\\
&=& \frac{|\Delta_0|^2}{g} - \sum_{s_2}\int_0^{\infty} \frac{k_{s_2} dk_{s_2}}{4\pi} \!\int_0^{2\pi}\! \frac{d\theta}{2\pi}
\left[\sqrt{\left(\frac{k_{s_2}^2}{2m} -\mu -\tilde{J} -\frac{\zeta^2}{2m} \right)^2 +|\Delta_{0}|^2} -\left|\frac{k_{s_2}^2}{2m} -\mu -\tilde{J} -\frac{\zeta^2}{2m} \right|\right]
\nonumber\\
&=& \frac{|\Delta_0|^2}{g} -\frac{\nu_0}{2} \sum_{s_2}\int_{-\omega_D}^{\omega_D} d\xi_{k,s_2} \int_0^{2\pi} \frac{d\theta}{2\pi}
\left(\sqrt{\xi_{k,s_2}^2 +|\Delta_{0}|^2} -|\xi_{k,s_2}|\right) \nonumber\\
&=&\frac{|\Delta_0|^2}{g} -\nu_0 \left[\omega_D\sqrt{\omega_D^2+|\Delta_0|^2} -\omega_D^2 +|\Delta_0|^2 \log{\left(\frac{\omega_D +\sqrt{\omega_D^2+|\Delta_0|^2}}{|\Delta_0|}\right)}\right] \nonumber\\
&\approx& \frac{|\Delta_0|^2}{g} -\nu_0\frac{|\Delta_0|^2}{2} \left[1 +2\ln{\left(\frac{2\omega_D}{|\Delta_0|}\right)}\right].
\end{eqnarray}

\begin{figure}[ht!]
\centering
\begin{subfloat}[]{\includegraphics[height=0.3\columnwidth]{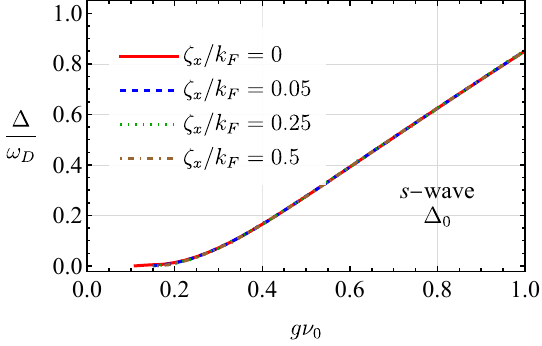}}
\end{subfloat}
\begin{subfloat}[]{\includegraphics[height=0.3\columnwidth]{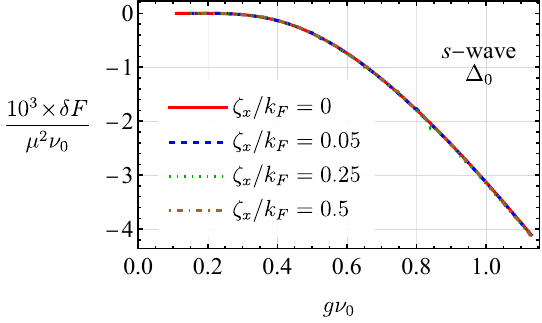}}
\end{subfloat}
\caption{\label{fig:2-model-gap-uniform-s-few-alpha}
The order parameter and the free energy difference for the uniform $s$-wave spin-singlet pairings $\hat{\Delta}(\mathbf{k}) = \Delta_0 (i\rho_y)$ for a few values of the spin splitting parameter $\zeta$. 
We use the first lines in Eqs.~\eqref{2-model-gap-eq-regular} and \eqref{2-model-F-S-N-2} supplemented with the requirement that only the energies in the interval $\left[-\omega_D,\omega_D\right]$ are relevant for pairing.
In all panels, we fix $T=10^{-5}\,\mu$, $t_{\rm inter} =\mu$, $\zeta_y=0$, and set the cutoff $\omega_D=0.1\,\mu$.
}
\end{figure}

To verify our result in Eq.~\eqref{2-model-gap-regular}, i.e., the independence of the gap on the separation vector, we present the numerical results for the gap and the free energy difference in Fig.~\ref{fig:2-model-gap-uniform-s-few-alpha}. Numerical results show a good agreement with our analytical calculations.

\subsubsection{$p$-wave equal-spin spin-triplet}
\label{sec:2-model-uniform-p}

In the case of $p$-wave gaps, $v(\mathbf{k})=\sqrt{2} \cos{\theta}$. The analysis of the gap equation and the free energy difference is similar to that in Sec.~\ref{sec:2-model-uniform-s}. We derive
\begin{eqnarray}
\label{2-model-gap-eq-Deltaz}
1 &\approx& g \frac{\nu_0}{4} \sum_{s_2=\pm} \int_{-\omega_D}^{\omega_D} d \xi_{s_2} \int_0^{2\pi} \frac{d\theta}{2\pi} \frac{\cos^2{\theta}}{\sqrt{\xi_{s_2}^2 +2\cos^2{\theta}|\Delta_{z}|^2}} = g \nu_0 \int_0^{2\pi} \frac{d\theta}{2\pi} \cos^2{\theta} \ln{\left(\frac{\sqrt{\omega_D^2 +2\cos^2{\theta}|\Delta_z|^2} +\omega_D}{\sqrt{\omega_D^2 +2\cos^2{\theta}|\Delta_z|^2} -\omega_D}\right)} \nonumber\\
&\approx& 2g \nu_0 \int_0^{2\pi} \frac{d\theta}{2\pi} \cos^2{\theta} \ln{\left(\frac{\sqrt{2}\omega_D}{|\Delta_z \cos{\theta}|}\right)} = g \nu_0\left[\ln{\left(\frac{2\omega_D}{|\Delta_z|}\right)} +\ln{\sqrt{2}} -1\right].
\end{eqnarray}
The above equation has the following solution:
\begin{equation}
\label{2-model-gap-Deltaz}
|\Delta_z| = 2\omega_D e^{-1/(g\nu_0) -(1-\ln{\sqrt{2}})}.
\end{equation}

The critical temperature is estimated in Eqs.~\eqref{2-model-gap-eq-regular-Tcr} and \eqref{2-model-gap-eq-regular-Tcr-fin}.

The free energy difference is 
\begin{eqnarray}
\label{2-model-F-S-N-Deltaz}
\delta F_z &=& \frac{|\Delta_z|^2}{g} -\nu_0 \int_0^{2\pi} \frac{d\theta}{2\pi} \Bigg[\omega_D\sqrt{\omega_D^2+2\cos^2{\theta}|\Delta_z|^2} -\omega_D^2 +
2\cos^2{\theta}|\Delta_z|^2 \log{\left(\frac{\omega_D +\sqrt{\omega_D^2 +2\cos^2{\theta}|\Delta_z|^2}}{\sqrt{2}|\cos{\theta}||\Delta_z|}\right)}\Bigg] \nonumber\\
&\approx& \frac{|\Delta_z|^2}{g} -|\Delta_z|^2 \int_0^{2\pi} \frac{d\theta}{2\pi} \cos^2{\theta} \left[1 +2\ln{\left(\frac{2\omega_D}{\sqrt{2}|\cos{\theta}||\Delta_z|}\right)}\right] = \frac{|\Delta_z|^2}{\tilde{g}} - \frac{|\Delta_z|^2}{2} \left[1 +2\ln{\left(\frac{2\omega_D}{|\Delta_z|}\right)} -(1-\ln{2})\right].\nonumber\\
\end{eqnarray}

\begin{figure}[ht!]
\centering
\begin{subfloat}[]{\includegraphics[height=0.3\columnwidth]{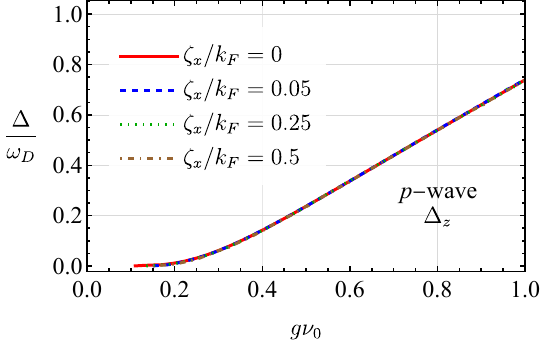}}
\end{subfloat}
\begin{subfloat}[]{\includegraphics[height=0.3\columnwidth]{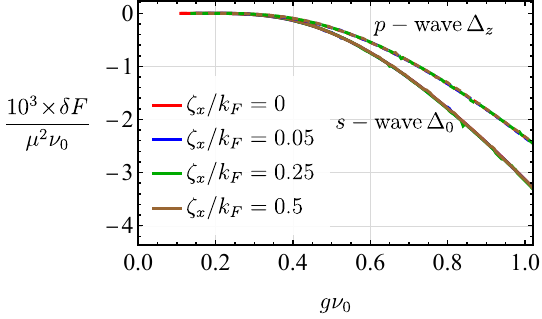}}\end{subfloat}
\caption{\label{fig:2-model-gap-uniform-p-few-alpha}
The order parameter and the free energy difference for the uniform $p$-wave mixed-spin spin-triplet pairings $\hat{\Delta}(\mathbf{k}) = \sqrt{2} \cos{(\theta)}\,\Delta_z\, i\rho_y \rho_z$ for a few values of the spin splitting parameter $\zeta_x$. 
We use the first lines in Eqs.~\eqref{2-model-gap-eq-regular} and \eqref{2-model-F-S-N-2} supplemented with the requirement that only the energies in the interval $\left[-\omega_D,\omega_D\right]$ are relevant for pairing; see also Fig.~\ref{fig:2-model-gap-uniform-s-few-alpha}(b) for the free energy for $s$-wave spin-singlet pairing.
In all panels, we fix $T=10^{-5}\,\mu$, $t_{\rm inter} =\mu$, $\zeta_y=0$, and set the cutoff $\omega_D=0.1\,\mu$.
}
\end{figure}

Using the solutions given in Eqs.~\eqref{2-model-gap-regular} and \eqref{2-model-gap-Deltaz}, we estimate the free energy differences as
\begin{eqnarray}
\label{2-model-F-S-N-Delta0-expl}
\delta F_0 &\approx& -\nu_0\frac{|\Delta_0|^2}{2} = -2\nu_0\omega_D^2e^{-2/(g\nu_0)},\\
\label{2-model-F-S-N-Deltaz-expl}
\delta F_z &\approx& -\nu_0|\Delta_z|^2 \left(1 - \frac{\ln{2}}{4}\right) =-\nu_0\left(1 - \frac{\ln{2}}{4}\right) e^{-2/(g\nu_0)} e^{-2(1-\ln{\sqrt{2}})}.
\end{eqnarray}
Numerically, $\delta F_0/\delta F_z \approx 2.23$. Hence, the $s$-wave pairing is more energetically favorable. This agrees with our numerical estimates in Fig.~\ref{fig:2-model-gap-uniform-p-few-alpha}(b).

\subsubsection{$s$-wave pairing in a full model}
\label{sec:2-model-uniform-s-full}

In this section, we present the results for the inter-band pairing in the full effective model \eqref{2-model-full}. We diagonalize the Hamiltonian \eqref{2-model-full} and consider pairing only between the bands at the Fermi level, i.e., we have the following BdG Hamiltonian:
\begin{equation}
\label{2-model-uniform-s-full-H-BdG}
\hat{H}_{\rm BdG} =
\begin{pmatrix}
H(\mathbf{k}) & \hat{\Delta} \\
\hat{\Delta}^{\dag} & -H^{*}(-\mathbf{k})
\end{pmatrix},
\end{equation}
where $H(\mathbf{k})$ in the diagonal basis is
\begin{eqnarray}
\label{2-model-uniform-s-full-H-BdG-1}
H(\mathbf{k}) &=& \mbox{diag}{\left\{\epsilon_j\right\}},\\
\label{2-model-uniform-s-full-H-BdG-2}
\hat{\Delta}(\mathbf{k}) &=& \frac{\rho_0+\rho_z}{2}\otimes (i \rho_y) \Delta_{12} + \frac{\rho_0-\rho_z}{2}\otimes (i \rho_y) \Delta_{34}
\end{eqnarray}
with the energies $\epsilon_j=\epsilon_{k,s_1,s_2}$ given in Eq.~\eqref{2-model-full-eps}.
Here, the constant inter-band pairing is consistent with the results of Sec.~\ref{sec:sm2-pairing-diag}.

\begin{figure}[ht!]
\centering
\begin{subfloat}[\label{fig:2-model-full-gap-12-a}]{\includegraphics[height=0.28\columnwidth]{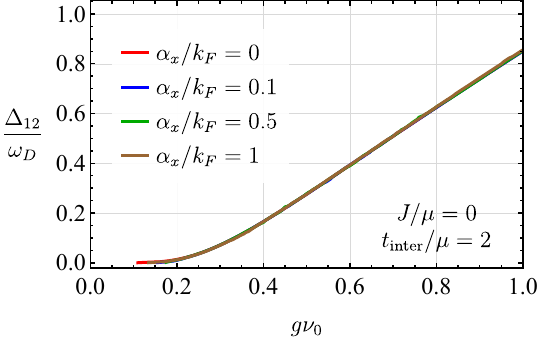}}
\end{subfloat}
\begin{subfloat}[\label{fig:2-model-full-gap-12-b}]{\includegraphics[height=0.28\columnwidth]{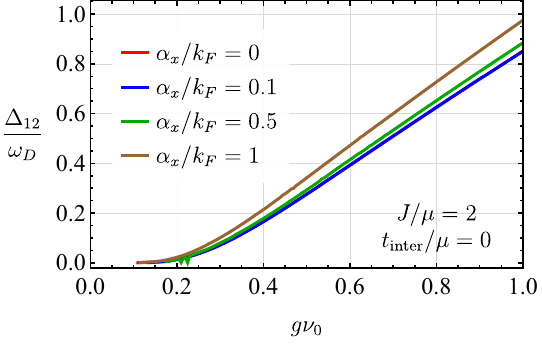}}
\end{subfloat}
\\
\begin{subfloat}[]{\includegraphics[height=0.28\columnwidth]{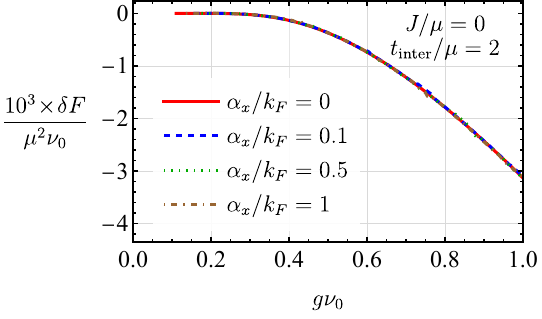}}
\end{subfloat}
\begin{subfloat}[]{\includegraphics[height=0.28\columnwidth]{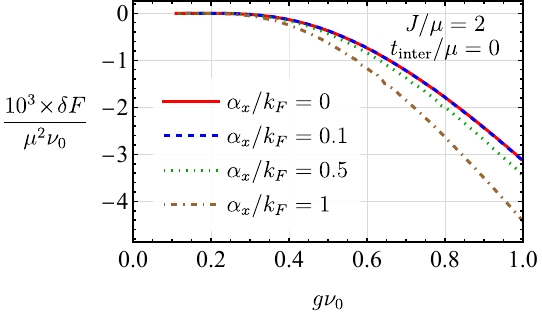}}
\end{subfloat}
\caption{\label{fig:2-model-full-gap-12}
The order parameter $\Delta_{12}$ and the free energy difference for the uniform $s$-wave inter-band pairings.
Top row: The magnitudes of the order parameter $\Delta_{12}$.
Bottom row: The difference of the free energies. 
In calculating the order parameter, only the energies in the interval $\left[-\omega_D,\omega_D\right]$ with respect to the Fermi level are relevant for pairing.
In all panels, we fix temperature $T=10^{-5}\,\mu$ and set the cutoff $\omega_D=0.1\,\mu$.
}
\end{figure}

We calculate the inter-band order parameter and the difference of the free energies in Fig.~\ref{fig:2-model-full-gap-12}. While there is a dependence on the spin-splitting parameter $\bm{\alpha}$, it is weak for well-separated bands and small values of $\bm{\alpha}$; these are the conditions for which our low-energy model \eqref{2-model-h-le} was derived.

Thus, the numerical results for the inter-band $s$-wave pairing in the full effective model suggest that the low-energy model \eqref{2-model-h-le} captures well the key aspects of the superconductivity in $p$M.

%% file: Supplemental-pMSC-3.tex
In this section, the details of the transport calculations are provided. We introduce the formalism for calculating the electric and spin conductances based on the scattering states and apply it to different types of junctions: normal metal (NM) with $p$M, \pmag with SC, and NM with a superconducting \pmag (SC$p$M).

To describe junctions with \pmag, we use the following Hamiltonian based on the low-energy model \eqref{2-model-h-le}:
\begin{eqnarray}
\label{lecm-a-ref-h-full}
H_{\rm BdG}(x) &=& \left[\frac{k_y^2-\nabla_x^2}{2m} -\mu -\tilde{J}(x)\right] \tau_z\otimes\rho_0 + \frac{\zeta_y(x) k_y}{m} \tau_0\otimes\rho_z \nonumber\\
&+&\frac{1}{2m}\left\{\zeta_x(x), -i\nabla_x\right\} \tau_0\otimes\rho_z
+U \df{x} \tau_z\otimes \rho_0 -\Delta(x) \tau_y \otimes\rho_y,
\end{eqnarray}
where $\bm{\tau}$ and $\bm{\rho}$ are Pauli matrices in the Nambu and band spaces, respectively, $\tilde{J}(x)=\sqrt{J^2(x) +t_{\rm inter}^2}$, 
$U$ is the potential barrier strength, $\left\{\ldots,\ldots\right\}$ is the anticommutator, and $\Delta$ is the $s$-wave singlet gap. 

This model describes a semi-infinite along the $x$-direction and infinite along the $y$-direction 2D junction. The configuration of the junctions is encoded in the coordinate-dependence of $\bm{\zeta}(x)$ and $\Delta(x)$. In the NM-$p$M junctions, we use $\tilde{J}(x)=t_{\rm inter}\Theta{(-x)} +\tilde{J} \Theta{(x)}$, 
$\bm{\zeta}(x)=\bm{\zeta} \Theta{(x)}$, and $\Delta(x)=0$ with $\Theta{(x)}$ being a unit step function. To model the $p$M-SC junction, we set $\tilde{J}(x)=\tilde{J} \Theta{(-x)} +t_{\rm inter}\Theta{(x)}$, $\bm{\zeta}(x)=\bm{\zeta} \Theta{(-x)}$ and $\Delta(x)=\Delta \Theta{(x)}$. Finally, in the NM-SC$p$M, we use $\tilde{J}(x)=t_{\rm inter}\Theta{(-x)} +\tilde{J} \Theta{(x)}$, $\bm{\zeta}(x)=\bm{\zeta} \Theta{(x)}$ and $\Delta(x)=\Delta \Theta{(x)}$.

\subsection{Scattering formalism}
\label{sec:sm3-model}

In the limit of vanishing temperature, the electric and spin conductances of a junction are defined as
\begin{eqnarray}
\label{lecm-a-G-el-all}
G_{{\rm el},j}(V) &=& e^2 \sum_{s}\int_{-\infty}^{\infty}\frac{dk_y}{(2\pi)^2} \left|\frac{\partial k_x}{\partial \epsilon}\right| \RE{\Psi_{s}^{\dag}(x) \hat{v}_j \tau_z\otimes \rho_0 \Psi_{s}(x)} \Big|_{\epsilon = eV},\\
\label{lecm-a-G-S-all}
G_{{\rm S},j}(V) &=& -e \sum_{s}\int_{-\infty}^{\infty}\frac{dk_y}{(2\pi)^2} \left|\frac{\partial k_x}{\partial \epsilon}\right| \RE{\Psi_{s}^{\dag}(x) \hat{v}_j \hat{s}_z \Psi_{s}(x)} \Big|_{\epsilon = eV},
\end{eqnarray}
where $j=x,y$ denotes the direction of the current and $V$ is the applied to the junction voltage bias. We sum over quasiparticle states $\Psi_{s}(x)$ of the impinging quasiparticles from the band $s$.
 
In the model \eqref{lecm-a-ref-h-full}, the velocity operators on the \pmag side are defined as
\begin{eqnarray}
\label{lecm-a-pM-vx}
\hat{v}_x &=& -i\nabla_x \tau_z\otimes\rho_0 + \zeta_x \tau_0\otimes\rho_z,\\
\label{lecm-a-pM-vy}
\hat{v}_y &=& k_y \tau_z\otimes \rho_0 + \zeta_y \tau_0\otimes \rho_z.
\end{eqnarray}
On the non-\pmag side, we set $\bm{\zeta}\to\mathbf{0}$.

The spin operator $\hat{s}_z$ in the band basis is given in Eq.~\eqref{2-model-sigmaz-le}. Therefore, we have the following expressions in the case of well-separated energy bands:
\begin{eqnarray}
\label{lecm-a-pM-vx-sigmaz-le}
\hat{v}_x \hat{s}_z &\approx& \frac{t_{\rm inter}}{\tilde{J}}\left[ -i  \nabla_x \tau_0\otimes\rho_z +\zeta_x \tau_z\otimes\rho_0 \right],\\
\label{lecm-a-pM-vy-sigmaz-le}
\hat{v}_y \hat{s}_z &\approx& \frac{t_{\rm inter}}{\tilde{J}}\left[ k_y \tau_0\otimes\rho_z +\zeta_y \tau_z\otimes\rho_0 \right].
\end{eqnarray}

The scattering states $\Psi_{s}(x)$ of the electrons belonging to the band $s$ have the following generic form at $x<0$:
\begin{eqnarray}
\label{lecm-a-ref-psi-e-x<0}
\Psi_{s}(x<0) &=&
\begin{pmatrix}
\psi_{s}\\
0\\
0
\end{pmatrix} e^{ik_{e,+,s}x} + \sum_{\sigma=\pm} a_{s,\sigma} \begin{pmatrix}
0\\
0\\
\psi_{\sigma}
\end{pmatrix} e^{ik_{h,+,\sigma}x}
+ \sum_{\sigma=\pm} b_{s,\sigma} \begin{pmatrix}
\psi_{\sigma}\\
0\\
0
\end{pmatrix} e^{ik_{e,-,\sigma}x}
\end{eqnarray}
and $x>0$:
\begin{eqnarray}
\label{lecm-a-ref-sc-psi-x>0}
\Psi_{s}(x>0) &=&
c_{s,+}\begin{pmatrix}
u\\
0\\
0\\
v
\end{pmatrix} e^{i p_{e,+,+}x}
+c_{s,-}\begin{pmatrix}
0\\
u\\
-v\\
0
\end{pmatrix} e^{ip_{e,+,-}x} +
d_{s,+}\begin{pmatrix}
0\\
v\\
-u\\
0
\end{pmatrix} e^{ip_{h,-,+}x}
+d_{s,-}\begin{pmatrix}
v\\
0\\
0\\
u
\end{pmatrix} e^{ip_{h,-,-}x}
\end{eqnarray}
with the standard coherence factors
\begin{eqnarray}
\label{lecm-a-ref-sc-uv}
u = \sqrt{\frac{\epsilon +\Omega}{2\epsilon}}, \quad \quad  v = \sqrt{\frac{\epsilon -\Omega}{2\epsilon}}
\end{eqnarray}
and
\begin{equation}
\label{lecm-NM-pMSC-Omega}
\Omega = \sign{\epsilon} \sqrt{\epsilon^2-\Delta^2}\, \Theta{\left(\epsilon^2-\Delta^2\right)} +i \sqrt{\Delta^2 -\epsilon^2}\, \Theta{\left(\Delta^2 -\epsilon^2\right)}.
\end{equation}
In addition, we used $\psi_{\sigma}$ defined in Eq.~\eqref{2-model-psi-le}.

The coefficients $a_{s,\pm}$, $b_{s,\pm}$, $c_{s,\pm}$, and $d_{s,\pm}$ follow from matching the wave functions at the interface $x=0$. The boundary conditions are obtained by integrating the eigenvalue equation $H_{\rm BdG}(x)\Psi(x)=\epsilon \Psi(x)$ in the vicinity of the interface, $\lim_{\delta x \to 0} \int_{-\delta x}^{\delta x} dx \left[H_{\rm BdG}(x)-\epsilon\right]\Psi(x)$. We derive:
\begin{eqnarray}
\label{lecm-a-ref-BC-1}
&&\partial_x \Psi(x>0) - \partial_x \Psi(x<0) = 2m U \tau_0\otimes\rho_0 \Psi(0) \pm i\zeta_x \tau_z\otimes \rho_z \Psi(0),\\
\label{lecm-a-ref-BC-2}
&&\Psi(x>0) =\Psi(x<0)=\Psi(0),
\end{eqnarray}
where the sign $\pm$ corresponds to the \pmag at $x<0$ ($+$) or $x>0$ ($-$).

Since there are no spin-fliping processes in the model, the expressions in Eqs.~\eqref{lecm-a-ref-psi-e-x<0} and \eqref{lecm-a-ref-sc-psi-x>0} can be simplified. We have $a_{+,+}=a_{-,-}=b_{+,-}=b_{-,+}=c_{+,-}=c_{-,+}=d_{+,+}=d_{-,-}=0$. This allows us to define $a_{s,\sigma}=a_{s} \delta_{s,-\sigma}$, $b_{s,\sigma}=b_{s} \delta_{s,\sigma}$, $c_{s,\sigma}=c_{s} \delta_{s,\sigma}$, and $d_{s,\sigma}=d_{s} \delta_{s,-\sigma}$. 

The wave vectors $k_{e/h,s_1,s_2}$ and $p_{e/h,s_1,s_2}$ depend on the type of the junction. In the NM-$p$M junction, we have
\begin{eqnarray}
\label{lecm-transverse-nmpm-kx}
k_{e,s_1,s_2} &=& s_1 \sqrt{2m\left(\mu +t_{\rm inter} +\epsilon\right) -k_y^2},\\
\label{lecm-transverse-nmpm-px}
p_{e,s_1,s_2} &=& -s_2 \zeta_x +s_1 \sqrt{2m\left(\mu +\tilde{J} +\epsilon\right) +\zeta^2 -(k_y+s_2\zeta_y)^2 },
\end{eqnarray}
where $\zeta=\sqrt{\zeta_x^2+\zeta_y^2}$.

In the $p$M-SC junction, the wave vectors are
\begin{eqnarray}
\label{lecm-transverse-pmsc-kx}
k_{e,s_1,s_2} &=& -s_2 \zeta_x +s_1 \sqrt{2m\left(\mu +\tilde{J} +\epsilon\right) +\zeta^2 -(k_y+s_2\zeta_y)^2},\\
\label{lecm-transverse-pmsc-px}
p_{e,s_1,s_2} &=& s_1 \sqrt{2m\left(\mu +t_{\rm inter} +\Omega\right) -k_y^2}.
\end{eqnarray}

Finally, in the NM-SC$p$M junction, we have
\begin{eqnarray}
\label{lecm-transverse-nmscpm-kx}
k_{e,s_1,s_2} &=& s_1 \sqrt{2m\left(\mu +t_{\rm inter} +\epsilon\right) -k_y^2},\\
\label{lecm-transverse-nmscpm-px}
p_{e,s_1,s_2} &=& -s_2 \zeta_x +s_1 \sqrt{2m\left(\mu +\tilde{J} +\Omega\right) +\zeta^2 -(k_y+s_2\zeta_y)^2}.
\end{eqnarray}
For holes, one has to replace $\epsilon \to -\epsilon$, $\Omega \to -\Omega$, and $s_2 \to -s_2$.

The Jacobian in Eqs.~\eqref{lecm-a-G-el-all} and \eqref{lecm-a-G-S-all}, i.e., $\partial k_x/(\partial \epsilon)$, acquires a simple form: $\left|\partial k_x/(\partial \epsilon)\right| = 1/(2|k_{e,s_1,s_2}+s_2\zeta_x|)$, where $\zeta_x$ should be set to zero if the impinging electrons correspond to the NM, i.e., in the NM-$p$M and NM-SC$p$M junctions.

Before presenting the details of the calculations and results for each of the junctions, let us discuss the coordinate dependence of the conductance. In agreement with the continuity equations, the longitudinal conductance does not depend on $x$. On the other hand, the transverse current can depend on $x$ due to surface-localized modes. Measuring such a coordinate dependence transport response requires a sufficiently fine spatial resolution $\sim 1/k_F$. Therefore, we consider spatially-averaged currents where the averaging is performed as $\aver{A} =\lim_{L_x\to\infty}\int_{-L_x}^{L_x}dx/(2L_x) A(x)$. The averaging allows us to disregard the contribution of the surface-localized modes and simplifies the interpretation of the results.

\subsection{NM-$p$M junction}
\label{sec:sm3-NMpM}

As a warmup, let us investigate the transport properties of the NM-$p$M junction. In our considerations, we use the model given in Eq.~\eqref{lecm-a-ref-h-full}, where, unlike the model used in Ref.~\cite{salehi_arxiv_24}, the spin-splitting parameter $\bm{\zeta}$ affects also the effective chemical potential on the \pmag side.

Since there is no superconductivity in the system, the scattering states \eqref{lecm-a-ref-psi-e-x<0} and \eqref{lecm-a-ref-sc-psi-x>0} can be significantly simplified: 
\begin{equation}
\label{lecm-transverse-psi-e-x<0}
\Psi_{s}(x<0) = \psi_s e^{ik_{e,+,s}x} + b_{s} \psi_s e^{ik_{e,-,s}x}
\end{equation}
on the NM side and
\begin{equation}
\label{lecm-transverse-psi-e-x>0}
\Psi_{s}(x>0) = c_{s} \psi_s e^{ip_{e,+,s}x},
\end{equation}
on the \pmag side. Here, $\psi_s$ is defined in Eq.~\eqref{2-model-psi-le}.

By using the boundary conditions \eqref{lecm-a-ref-BC-1} and \eqref{lecm-a-ref-BC-2}, we obtain the following coefficients:
\begin{equation}
\label{lecm-transverse-b-c}
b_{s} = c_{s} -1, \quad \quad
c_{s} = \frac{2k_{e,+,+}}{k_{e,+,+}+p_{s} +ik_FZ},
\end{equation}
where we used Eqs.~\eqref{lecm-transverse-nmpm-kx} and \eqref{lecm-transverse-nmpm-px} as well as defined shorthands $p_{s} = p_{e,+,s}+s\zeta_x= \sqrt{2m(\mu+\tilde{J} +\epsilon)+\zeta^2 -(k_y+s\zeta_y)^2}$ and $Z=2mU/k_F$.

The electric current densities are
\begin{eqnarray}
\label{lecm-transverse-jx-nm}
j_{{\rm el}, x}^{\rm (NM)} &=& \RE{\Psi_{s}^{\dag}(x<0) \hat{v}_x \Psi_{s}(x<0)} = k_{e,+,s} \left(1 - |b_s|^2\right) = \frac{4 p_s k_{e,+,+}^2}{\left(k_{e,+,+}+p_{s}\right)^2+(k_FZ)^2},\\
\label{lecm-transverse-jy-nm}
j_{{\rm el}, y}^{\rm (NM)} &=& \RE{\Psi_{s}^{\dag}(x<0) \hat{v}_y \Psi_{s}(x<0)} = k_y \left(1 + |b_s|^2\right) = \frac{4 k_y p_sk_{e,+,+}}{\left(k_{e,+,+}+p_{s}\right)^2+(k_FZ)^2},\\
\label{lecm-transverse-jx-pm}
j_{{\rm el}, x}^{(p{\rm M})} &=& \RE{\Psi_{s}^{\dag}(x>0) \hat{v}_x \Psi_{s}(x>0)} = p_s |c_s|^2 = \frac{4p_sk_{e,+,+}^2}{(k_{e,+,+}+p_{s})^2 +(k_FZ)^2},\\
\label{lecm-transverse-jy-pm}
j_{{\rm el}, y}^{(p{\rm M})} &=& \RE{\Psi_{s}^{\dag}(x>0) \hat{v}_y \Psi_{s}(x>0)} = (k_y+s\zeta_y) |c_s|^2 = \frac{4(k_y+s\zeta_y)k_{e,+,+}^2}{(k_{e,+,+}+p_{s})^2 +(k_FZ)^2}.
\end{eqnarray}
In the spin currents, one should multiply the above expressions by $s$. In view of the structure of $j_{{\rm el}, y}^{\rm (NM)}$ and $j_{{\rm el}, y}^{(p{\rm M})}$, the corresponding electric conductances vanish $G_{{\rm el}, y}=0$ after integration over $k_y$ even if $\zeta_y\neq 0$ since the corresponding current is antisymmetric with respect to $k_y \to -k_y$ and $s\to -s$. A similar symmetry argument can be applied to the longitudinal spin conductance, i.e., $G_{{\rm S}, x}=0$. The transverse spin conductance, on the other hand, is nonvanishing at $\zeta_y\neq0$ because the corresponding current is even with respect to $k_y \to -k_y$ and $s\to -s$. Finally, we note that, in calculating the conductances, we take into account only propagating modes, i.e., the modes with real wave vectors. Impinging waves are always propagating with $\IM{k_{e,+,s}}=0$.

\begin{figure}[!ht]
\centering
\begin{subfloat}[]{\includegraphics[width=0.32\columnwidth]{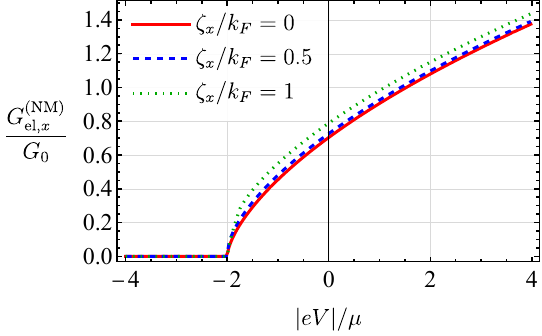}}\end{subfloat}
\begin{subfloat}[]{\includegraphics[width=0.32\columnwidth]{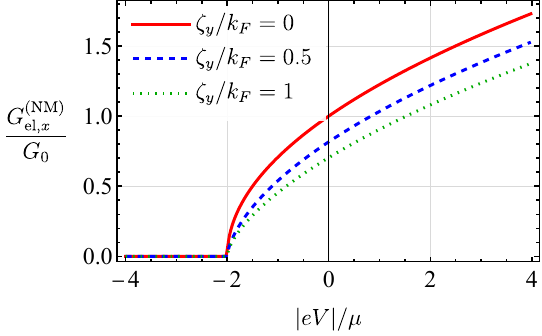}}\end{subfloat}
\begin{subfloat}[]{\includegraphics[width=0.32\columnwidth]{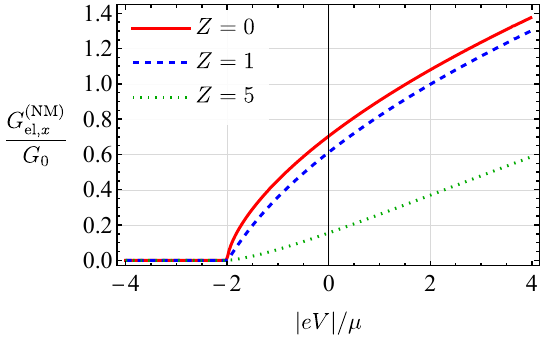}}\end{subfloat}
\\
\begin{subfloat}[]{\includegraphics[width=0.32\columnwidth]{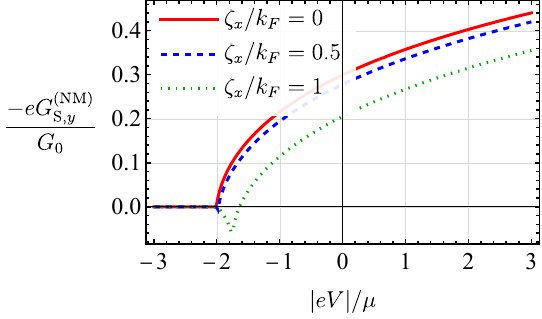}}\end{subfloat}
\begin{subfloat}[]{\includegraphics[width=0.32\columnwidth]{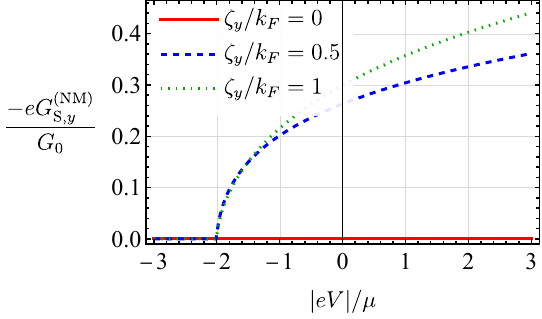}}\end{subfloat}
\begin{subfloat}[]{\includegraphics[width=0.32\columnwidth]{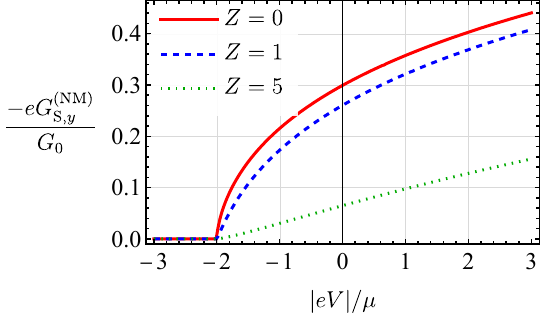}}\end{subfloat}
\caption{\label{fig:sm3-NMpM-G}
The longitudinal electric (top row) and transverse spin (bottom row) conductance of the NM-$p$M junction for different combinations of parameters as a function of the voltage bias; see Eqs.~(\ref{lecm-a-G-el-all}) and (\ref{lecm-a-G-S-all}) for the definitions.
We normalize by longitudinal electric conductance $G_{0}=G_{{\rm el},x}(0)$ calculated at $eV = 0$, $t_{\rm inter}=\mu$, and $Z=J=\zeta=0$. 
In all panels, unless otherwise stated, $t_{\rm inter}=\mu$, $\zeta_x=0$, and $\zeta_y=k_F$.
}
\end{figure}

We show the longitudinal electric and transverse spin conductance in Fig.~\ref{fig:sm3-NMpM-G} for a few values of the spin-splitting parameters $\zeta_x$ and $\zeta_y$ as well as the barrier height $Z$. The dependence on $\zeta_x$ and $J$ is similar and originates from the mismatch of the effective chemical potentials on the NM and \pmag sides of the junction, see the $\sim \zeta^2$ term in Eq.~\eqref{lecm-transverse-nmpm-px}. The splitting along the junction $\zeta_y$ reduces the overlap of the Fermi surfaces on both sides of the junction leading to a lower electric conductance, see Fig.~\ref{fig:sm3-NMpM-G}(b). As expected, the potential barrier quantified by $Z$ decreases the conductance, see Fig.~\ref{fig:sm3-NMpM-G}(c).

For the transverse spin conductance $G_{{\rm S}, y}$, the spin splitting $\zeta_y$ is crucial. As with the electric conductance, the spin conductance depends on the overlap of the Fermi surfaces, hence the growth with $\zeta_y$ is different at different $eV$. For both electric and spin transport, the conductance vanishes for $eV<-|\mu+t_{\rm inter}|$.

\subsection{$p$M-SC junction}
\label{sec:sm3-pMSC}

In this section, we discuss the conductances in the $p$M-SC junction.

We start with presenting the expressions for the different components of currents. The longitudinal electric and spin currents do not depend on $x$ on both sides of the junction. For example, we have the following electric currents:
\begin{eqnarray}
\label{lecm-a-ref-jxElpM}
j_{{\rm el},x}(x<0) &=& \RE{\Psi_{s}^{\dag}(x<0) \hat{v}_x \tau_z \otimes \rho_0 \Psi_{s}(x<0)} \nonumber\\
&=& \left(k_{e,+,s} +s \zeta_x \right) +|a_{s}|^2 \left(k_{h,+, -s} +s \zeta_x \right) \Theta{\left(-\left|\IM{k_{h,+,-s}}\right|\right)} +|b_{s}|^2 \left(k_{e,-,s} +s \zeta_x \right) \Theta{\left(-\left|\IM{k_{e,-,s}}\right|\right)},\nonumber\\
\\
\label{lecm-a-ref-jxElSC}
j_{{\rm el},x}(x>0) &=& \RE{\Psi_{s}^{\dag}(x>0) \hat{v}_x \tau_z \otimes \rho_0 \Psi_{s}(x>0)} \nonumber\\
&=& (|u|^2+|v|^2) \left[|c_{s}|^2 p_{e,+,s} \Theta{\left(-\left|\IM{p_{e,+,s}}\right|\right)} +|d_{s}|^2 p_{h,-,-s} \Theta{\left(-\left|\IM{p_{h,-,-s}}\right|\right)} \right],
\end{eqnarray}
where the $\Theta$-functions denote the fact that only propagating modes contribute to the longitudinal current and there are no quasiparticle currents for subgap energies. Furthermore, here and in what follows, the impinging waves are always propagating with $\IM{k_{e,+,s}}=0$.

The spin currents are
\begin{eqnarray}
\label{lecm-a-ref-jxSpM}
j_{{\rm S},x}(x<0) &=& \RE{\Psi_{s}^{\dag}(x<0) \hat{v}_x \hat{s}_z \Psi_{s}(x<0)} \nonumber\\ 
&=& s\frac{t_{\rm inter}}{\tilde{J}} \Bigg[\left(k_{e,+,s} +s\zeta_x \right) - |a_{s}|^2 \left(k_{h,+,-s} +s \zeta_x \right)  \Theta{\left(-\left|\IM{k_{h,+,-s}}\right|\right)} \nonumber\\
&+& |b_{s}|^2 \left(k_{e,-,s} +s \zeta_x \right) \Theta{\left(-\left|\IM{k_{e,-,s}}\right|\right)} \Bigg],\\
\label{lecm-a-ref-jxSSC}
j_{{\rm S},x}(x>0) &=& \RE{\Psi_{s}^{\dag}(x>0) \hat{v}_x \hat{s}_z \Psi_{s}(x>0)} \nonumber\\  
&=& s(|u|^2-|v|^2) \left[ p_{e,+,s} |c_{s}|^2 \Theta{\left(-\left|\IM{p_{e,+,s}}\right|\right)} - p_{h,-,-s} |d_{s}|^2 \Theta{\left(-\left|\IM{p_{h,-,-s}}\right|\right)}\right].\nonumber\\
\end{eqnarray}

We note that while the longitudinal conductance does not depend on $x$, the transverse spin conductance may depend on the coordinate. Indeed, we have the following expressions:
\begin{eqnarray}
\label{lecm-a-ref-jyElpM}
j_{{\rm el},y}(x<0) &=& \RE{\Psi_{s}^{\dag}(x<0) \hat{v}_y \tau_z \otimes \rho_0 \Psi_{s}(x<0)} \nonumber\\
&=& \left(k_y +s \zeta_y \right) \left[1 + |a_{s}|^2 e^{-2\IM{k_{h,+,-s}} x}
+|b_{s}|^2 e^{-2\IM{k_{e,-,s}} x} \right] +2\left(k_y +s \zeta_y \right) \RE{b_{s} e^{i(k_{e,-,s} -k_{e,+,s})x}},\nonumber\\
\\
\label{lecm-a-ref-jyElSC}
j_{{\rm el},y}(x>0) &=& \RE{\Psi_{s}^{\dag}(x>0) \hat{v}_y \tau_z \otimes \rho_0 \Psi_{s}(x>0)} \nonumber\\
&=& k_y (|u|^2+|v|^2)\left[ |c_{s}|^2  e^{i(p_{e,+,s}-p_{e,+,s}^*)x} + |d_{s}|^2 e^{i(p_{h,-,-s}-p_{h,-,-s}^*)x}\right] \nonumber\\
&+& 2k_y (v^*u+u^*v) \RE{c^*_{s}d_{s} e^{i(p_{h,-,-s}-p_{e,+,s}^*)x}},\\
\label{lecm-att-ref-jySpM}
j_{{\rm S},y}(x<0) &=& \RE{\Psi_{s}^{\dag}(x<0) \hat{v}_y \hat{s}_z \Psi_{s}(x<0)} \nonumber\\ 
&=& s \frac{t_{\rm inter}}{\tilde{J}} \left(k_y +s \zeta_y \right) \left[
1 - |a_{s}|^2 e^{-2\IM{k_{h,+,-s}} x}  +|b_{s}|^2 e^{-2\IM{k_{e,-,s}} x} \right] \nonumber\\
&+& 2 s \left(k_y +s \zeta_y \right) \RE{b_{s} e^{i(k_{e,-,s} -k_{e,+,s})x}},\\
\label{lecm-a-ref-jySSC}
j_{{\rm S},y}(x>0) &=& \RE{\Psi_{s}^{\dag}(x>0) \hat{v}_y \hat{s}_z \Psi_{s}(x>0)} \nonumber\\
&=& s k_y (|u|^2-|v|^2)\left[ |c_{s}|^2 e^{i(p_{e,+,s}-p_{e,+,s}^*)x} - |d_{s}|^2 e^{i(p_{h,-,-s}-p_{h,-,-s}^*)x}\right] \nonumber\\
&+& 2 s k_y (v^*u+u^*v) \RE{c^*_{s}d_{s} e^{i(p_{h,-,-s}-p_{e,+,s}^*)x}}.
\end{eqnarray}
As follows from the above expressions, transverse currents depend on the $x$-coordinate due to modes localized at the interface. By performing averaging $\aver{A} =\lim_{L_x\to\infty}\int_{-L_x}^{L_x}dx/(2L_x) A(x)$, such contributions become negligible:
\begin{eqnarray}
\label{lecm-a-ref-jyElpM-aver}
\aver{j_{{\rm el},y}(x<0)} &=& \left(k_y +s \zeta_y \right) \left[1 + |a_{s}|^2 \Theta{\left(-\left|\IM{k_{h,+,-s}}\right|\right)} +|b_{s}|^2 \Theta{\left(-\left|\IM{k_{e,-,s}}\right|\right)} \right],\\
\label{lecm-a-ref-jyElSC-aver}
\aver{j_{{\rm el},y}(x>0)} &=&  k_y (|u|^2+|v|^2)\left[ |c_{s}|^2  \Theta{\left(-\left|\IM{p_{e,+,s}}\right|\right)} + |d_{s}|^2 \Theta{\left(-\left|\IM{k_{h,-,-s}}\right|\right)} \right],\\
\label{lecm-att-ref-jySpM-aver}
\aver{j_{{\rm S},y}(x<0)} &=& s \frac{t_{\rm inter}}{\tilde{J}} \left(k_y +s \zeta_y \right) \left[1 - |a_{s}|^2 \Theta{\left(-\left|\IM{k_{h,+,-s}}\right|\right)} + |b_{s}|^2 \Theta{\left(-\left|\IM{k_{e,-,s}}\right|\right)} \right],\\
\label{lecm-a-ref-jySSC-aver}
\aver{j_{{\rm S},y}(x>0)} &=& s k_y (|u|^2-|v|^2)\left[ |c_{s}|^2 \Theta{\left(-\left|\IM{p_{e,+,s}}\right|\right)} - |d_{s}|^2 \Theta{\left(-\left|\IM{k_{h,-,-s}}\right|\right)}\right].
\end{eqnarray}

We use expressions in Eqs.~\eqref{lecm-a-ref-jxElpM}--\eqref{lecm-a-ref-jxSSC} and Eqs.~\eqref{lecm-a-ref-jyElpM-aver}--\eqref{lecm-a-ref-jySSC-aver} in calculating the electric \eqref{lecm-a-G-el-all} and spin \eqref{lecm-a-G-S-all} conductances. Since the corresponding currents are antisymmetric with respect to $k_y \to -k_y$ and $s \to -s$, the longitudinal spin and transverse electric conductances in the $p$M-SC junction vanish.

Before presenting the numerical results for the conductances, let us consider the transport coefficients $a_{s}$, $b_{s}$, $c_s$, and $d_s$. In a general case, the corresponding expressions are cumbersome but can be straightforwardly obtained. We notice that the Andreev reflection amplitude $a_{s} \propto p_{e,+,s} -p_{h,-,s}$. Therefore, it vanishes if $p_{e,+,s}$ and $p_{h,-,s}$ are purely imaginary, which is the case when matching of the transverse wave vectors is impossible. Then, $|b_{s}|=1$ leading to total normal reflection. Therefore, in our analytical considerations, we focus on the case where all modes are propagating, i.e., the corresponding wave vectors are real. To further simplify our formulas, we consider the Andreev approximation where $|\epsilon|,|\Delta| \ll \mu$. 
Then, we have
\begin{eqnarray}
\label{lecm-andreev-2-a}
a_{s} &=& \frac{\sign{\epsilon} |\Delta|}{\Omega} \frac{2s p_{e,+,+} k_{s}}{(p_{e,+,+}+k_{s})^2 -2p_{e,+,+} k_{s}\left(1 - \epsilon/\Omega\right) +(k_FZ)^2},\\
\label{lecm-andreev-2-b}
b_{s} &=& - \frac{p_{e,+,+}^2 -k_{s}^2 +k_FZ\left(k_FZ +2ik_{s}\right)}{(p_{e,+,+}+k_{s})^2 -2p_{e,+,+} k_{s}\left(1 - \epsilon/\Omega\right) +(k_FZ)^2},\\
\label{lecm-andreev-2-c}
c_{s} &=& \frac{\epsilon}{\Omega} \sqrt{\frac{\epsilon +\Omega}{2\epsilon}} \frac{2k_s\left(p_{e,+,+} +k_s -ik_FZ\right)}{(p_{e,+,+}+k_{s})^2 -2p_{e,+,+} k_{s}\left(1 - \epsilon/\Omega\right) +(k_FZ)^2},\\
\label{lecm-andreev-2-d}
d_{s} &=& \frac{\epsilon}{\Omega} \sqrt{\frac{\epsilon -\Omega}{2\epsilon}} \frac{2k_s\left(p_{e,+,+} -k_s +ik_FZ\right)}{(p_{e,+,+}+k_{s})^2 -2p_{e,+,+} k_{s}\left(1 - \epsilon/\Omega\right) +(k_FZ)^2},
\end{eqnarray}
where $p_{e,+,+} = \sqrt{2m(\mu+t_{\rm inter}) -k_y^2}$ and $k_{s} = \sqrt{2m(\mu+\tilde{J})+\zeta^2 -(k_y+s\zeta_y)^2}$.

In the case of a tunneling junction with $Z\to \infty$,
\begin{align}
\label{lecm-andreev-2-a-b-Z-inf}
&\lim_{Z\to \infty} a_{s} \approx \frac{\sign{\epsilon} |\Delta|}{\Omega} \frac{2s  p_{e,+,+} k_{s}}{(k_FZ)^2}, &\quad \quad 
&\lim_{Z\to \infty} b_{s} \approx  -1 -2i \frac{k_{s}}{Z} +\frac{2k_{s}}{(k_FZ)^2} \left(k_{s} +\frac{\epsilon p_{e,+,+}}{\Omega}\right),\\
\label{lecm-andreev-2-c-d-Z-inf}
&\lim_{Z\to \infty} c_{s} \approx -i\frac{\epsilon}{\Omega} \sqrt{\frac{\epsilon +\Omega}{2\epsilon}} \frac{2k_s}{k_FZ}, &\quad \quad
&\lim_{Z\to \infty} d_{s} \approx i\frac{\epsilon}{\Omega} \sqrt{\frac{\epsilon -\Omega}{2\epsilon}} \frac{2k_s}{k_FZ}.
\end{align}
Then, the corresponding scattering amplitudes are
\begin{eqnarray}
\label{lecm-andreev-2-a2-Z-inf}
\lim_{Z\to \infty} |a_s|^2 &\approx& \frac{4|\Delta|^2}{|\epsilon^2 -\Delta^2|} \frac{p_{e,+,+}^2 k_{s}^2}{(k_FZ)^4},\\
\label{lecm-andreev-2-b2-Z-inf}
\lim_{Z\to \infty} |b_s|^2 &\approx&  1 -\frac{2\epsilon}{\sqrt{|\epsilon^2 -\Delta^2|}} \frac{p_{e,+,+}k_{s}}{(k_FZ)^2} \Theta{(|\epsilon|-|\Delta|)}
+\frac{8\epsilon}{\sqrt{|\epsilon^2 -\Delta^2|}} \frac{p_{e,+,+} k_{s}^2}{(k_FZ)^3}\Theta{(|\Delta| -|\epsilon|)},\\
\label{lecm-andreev-2-c2-Z-inf}
\lim_{Z\to \infty} |c_s|^2 &\approx& \frac{\left|\epsilon \left(\epsilon - \sqrt{\epsilon^2 -\Delta^2}\right)\right|}{|\epsilon^2 -\Delta^2|} \frac{2 k_s^2}{(k_FZ)^2} \Theta{(|\epsilon|-|\Delta|)} +\frac{|\epsilon \Delta|}{|\epsilon^2 -\Delta^2|} \frac{2 k_s^2}{(k_FZ)^2} \Theta{(|\Delta|-|\epsilon|)},\\
\label{lecm-andreev-2-d2-Z-inf}
\lim_{Z\to \infty} |d_s|^2 &\approx&  \frac{\left|\epsilon \left(\epsilon + \sqrt{\epsilon^2 -\Delta^2}\right)\right|}{|\epsilon^2 -\Delta^2|} \frac{2 k_s^2}{(k_FZ)^2} \Theta{(|\epsilon|-|\Delta|)} +\frac{|\epsilon \Delta|}{|\epsilon^2 -\Delta^2|} \frac{2 k_s^2}{(k_FZ)^2} \Theta{(|\Delta|-|\epsilon|)}.
\end{eqnarray}
As expected, the Andreev retroreflection is suppressed compared to the normal reflection. The latter is significant for overgap energies and has $\sim 1/\sqrt{|\epsilon^2 -\Delta^2|}$ divergence at $|\epsilon| \searrow |\Delta|$. 

\begin{comment}
\begin{figure}[ht]
\centering
\begin{subfloat}[]{\includegraphics[width=0.45\columnwidth]{SM-Coefficients-pMSC-eps=0-mu=1000.-alphaX=0-alphaY=2.-Z=0-no-ev.pdf}}
\end{subfloat}
\begin{subfloat}[]{\includegraphics[width=0.45\columnwidth]{SM-Energy-pMSC-mu=1000.-alphaX=0-alphaY=2.-Z=0-no-ev.pdf}}
\end{subfloat}
\caption{\label{fig:lecm-a-2-coefficients}
Panel (a): Andreev retroreflection $|a_{s}|^2$ and normal-reflection $|b_{s}|^2$ transport coefficients as functions of $k_y$. In plotting the coefficients, we took into account that the corresponding modes should be propagating and set $\epsilon=0$.
Panel (b): The energy spectrum showing the spin-up (solid red) and spin-down (dashed blue) energy bands of the $p$M. The normal-state energy spectrum in the SC is shown by the dotted black line.
In all panels, we used $t_{\rm inter}=\mu$, $Z=J=0$, $\zeta_x=0$, and $\zeta_y=k_F$.
}
\end{figure}
\end{comment}

\begin{figure}[ht]
\centering
\begin{subfloat}[]{\includegraphics[width=0.31\columnwidth]{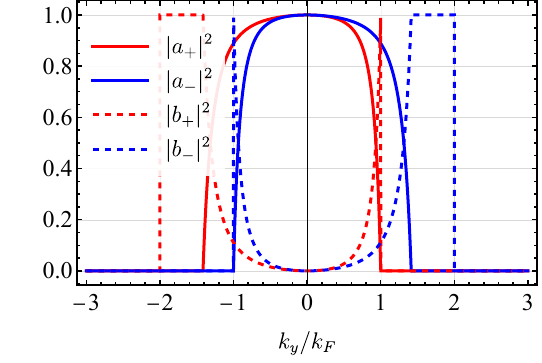}}
\end{subfloat}
\begin{subfloat}[]{\includegraphics[width=0.31\columnwidth]{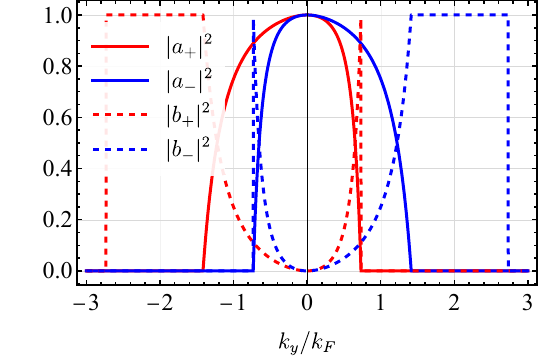}}
\end{subfloat}
\begin{subfloat}[]{\includegraphics[width=0.31\columnwidth]{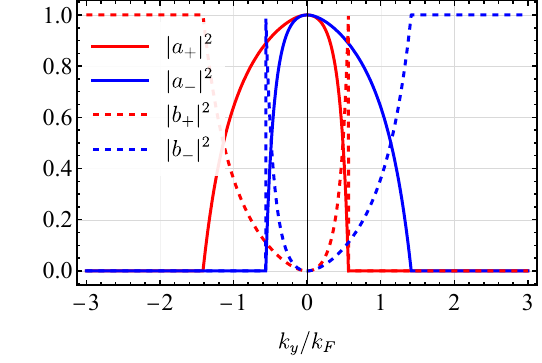}}
\end{subfloat}\\
\begin{subfloat}[]{\includegraphics[width=0.31\columnwidth]{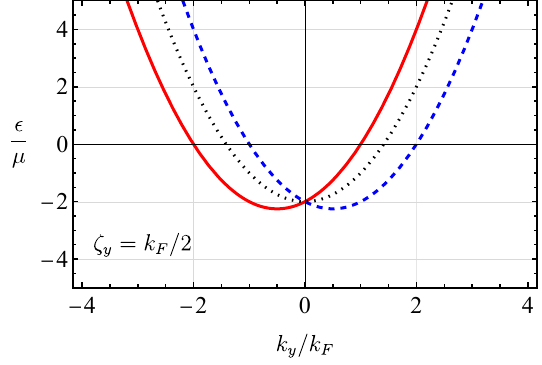}}
\end{subfloat}
\begin{subfloat}[]{\includegraphics[width=0.31\columnwidth]{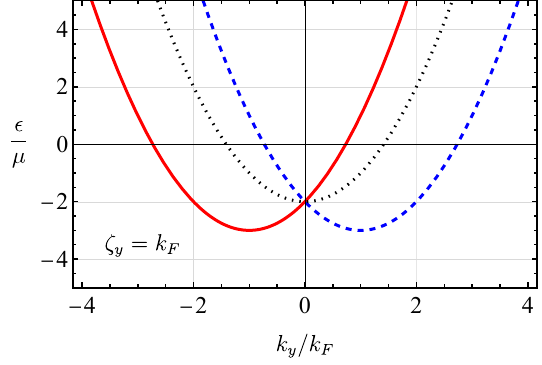}}
\end{subfloat}
\begin{subfloat}[]{\includegraphics[width=0.31\columnwidth]{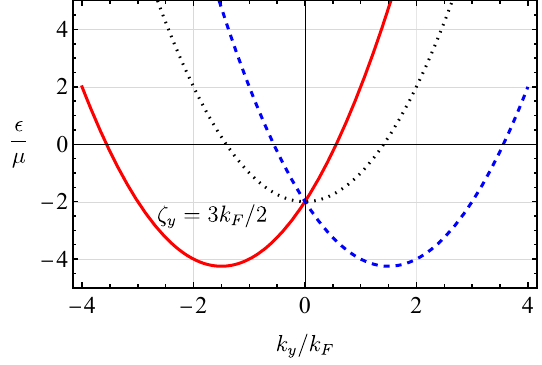}}
\end{subfloat}
\caption{\label{fig:lecm-a-2-coefficients}
Top row: Andreev retroreflection $|a_{s}|^2$ and normal-reflection $|b_{s}|^2$ transport coefficients as functions of $k_y$. In plotting the coefficients, we took into account that the corresponding modes should be propagating and set $\epsilon=0$.
Bottom row: The energy spectrum showing the spin-up (solid red) and spin-down (dashed blue) energy bands of the $p$M. The normal-state energy spectrum in the SC is shown by the dotted black line. The spin-splitting parameter $\zeta_y$ is $\zeta_y=k_F/2$ (left column), $\zeta_y=k_F$ (middle column), and $\zeta_y=3k_F/2$ (right column). In all panels, we use $t_{\rm inter}=\mu$, $Z=J=0$, and $\zeta_x=0$.
}
\end{figure}

We visualize the transport coefficients on the non-superconducting side for a few values of $\zeta_y$ in Fig.~\ref{fig:lecm-a-2-coefficients}. The Andreev retroreflection coefficient $|a_{s}|^2$ requires an overlap of the normal-state FS in the SC (dotted black line) in the bottom row of Fig.~\ref{fig:lecm-a-2-coefficients} and the FS in the $p$M (red and blue lines). Normal reflection is complete with $|b_{s}|^2=1$ when the FS in $p$M has no overlap with its counterpart on the SC side and decreases if such overlap is possible. The normal reflection coefficient $b_s$ vanishes for momenta outside the FS in $p$M since there are no propagating quasiparticles in this case.

\begin{figure}[ht!]
\centering
\begin{subfloat}[]{\includegraphics[width=0.31\columnwidth]{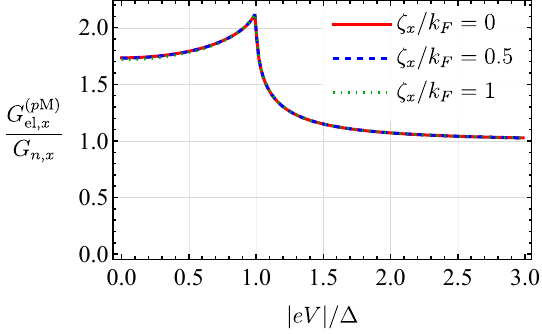}}\end{subfloat}
\begin{subfloat}[]{\includegraphics[width=0.31\columnwidth]{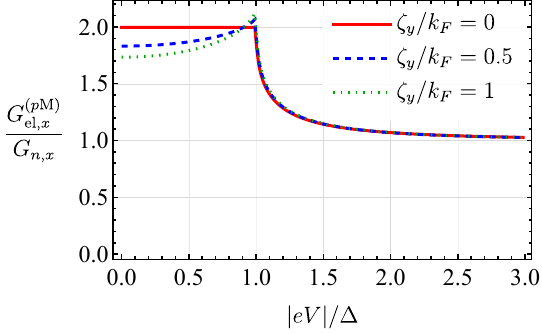}}\end{subfloat}
\begin{subfloat}[]{\includegraphics[width=0.31\columnwidth]{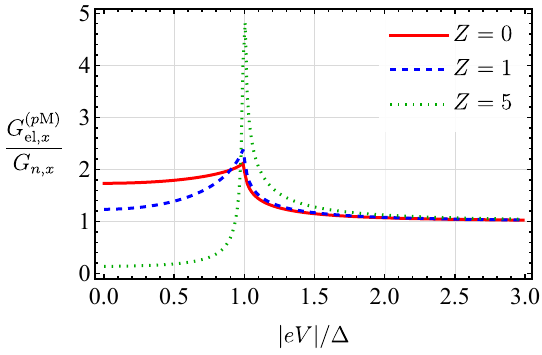}}\end{subfloat}
\\
\begin{subfloat}[]{\includegraphics[width=0.31\columnwidth]{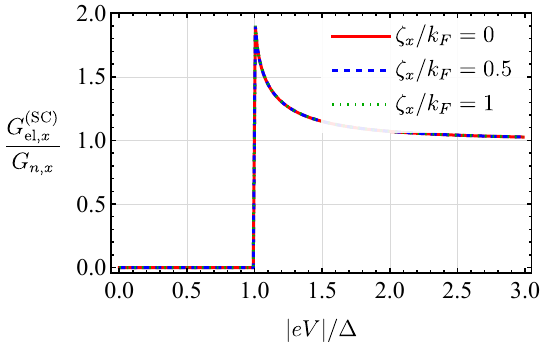}}\end{subfloat}
\begin{subfloat}[]{\includegraphics[width=0.31\columnwidth]{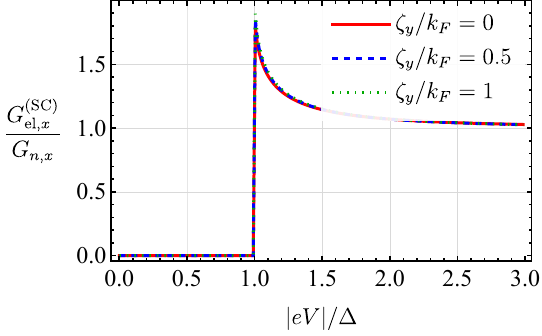}}\end{subfloat}
\begin{subfloat}[]{\includegraphics[width=0.31\columnwidth]{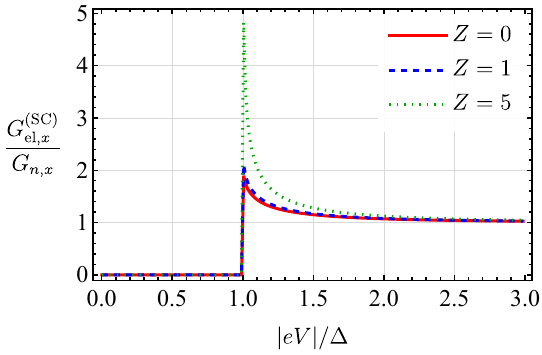}}\end{subfloat}
\caption{\label{fig:sm3-pMSC-G-el}
The longitudinal electric conductance $G_{{\rm el}, x}$ for different combinations of parameters as a function of the voltage bias; see Eq.~\eqref{lecm-a-G-el-all} for the definition.
The top and bottom rows show the conductance on the \pmag and SC sides, respectively.
We normalize by the longitudinal normal-state electric conductance $G_{n,x}$ calculated as $G_{{\rm el},x}^{\rm (p{\rm M})}$ at $eV = 10\,\Delta$.
In all panels, unless otherwise stated, $t_{\rm inter}=\mu$, $\zeta_x=0$, $\zeta_y=k_F$, $J=Z=0$, and $\mu=10^3\,\Delta$.
}
\end{figure}

We show the longitudinal electric conductance $G_{{\rm el},x}$ in Fig.~\ref{fig:sm3-pMSC-G-el}. As in regular NM-SC junctions, there is an Andreev plateau where the conductance is twice as high as in the normal state. The depression at $eV\to0$ in Fig.~\ref{fig:sm3-pMSC-G-el}(a) is caused by the mismatch of the effective chemical potential at $\zeta_y\neq0$. Since the electric conductance is determined by the overlap of the two Fermi surfaces, which is affected by $\zeta_y$ but not $\zeta_x$, there is a noticeable difference in the conductances in Figs.~\ref{fig:sm3-pMSC-G-el}(a) and \ref{fig:sm3-pMSC-G-el}(b).
Potential barrier quantified by $Z$ suppresses the Andreev retroreflection compared to the normal reflection, see also Eqs.~\eqref{lecm-andreev-2-a2-Z-inf} and \eqref{lecm-andreev-2-b2-Z-inf} for the corresponding transport coefficients, leading to a well-pronounced peak at $|eV| \searrow \Delta$, see Fig.~\ref{fig:sm3-pMSC-G-el}(c). On the SC side, the longitudinal electric conductance for subgap energies vanishes because there are no propagating quasiparticles. 

\begin{figure}[ht!]
\centering
\begin{subfloat}[]{\includegraphics[width=0.31\columnwidth]{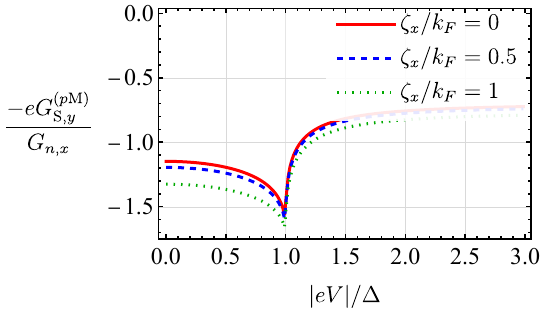}}\end{subfloat}
\begin{subfloat}[]{\includegraphics[width=0.31\columnwidth]{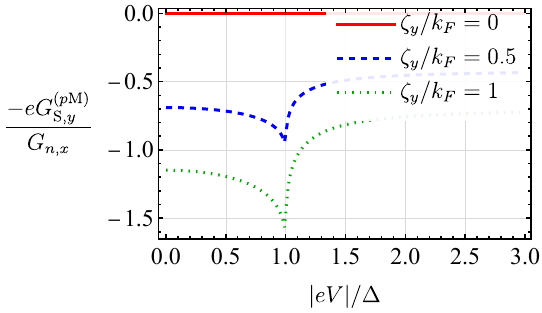}}\end{subfloat}
\begin{subfloat}[]{\includegraphics[width=0.31\columnwidth]{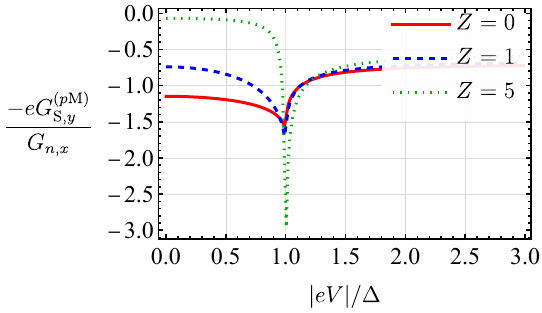}}\end{subfloat}
\\
\begin{subfloat}[]{\includegraphics[width=0.31\columnwidth]{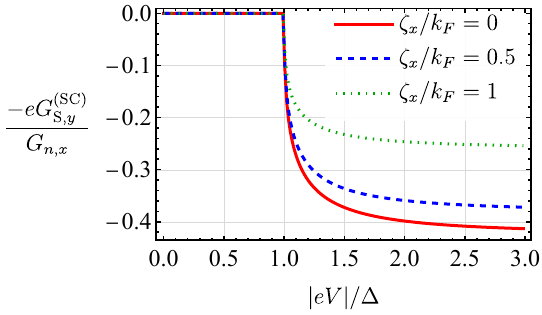}}\end{subfloat}
\begin{subfloat}[]{\includegraphics[width=0.31\columnwidth]{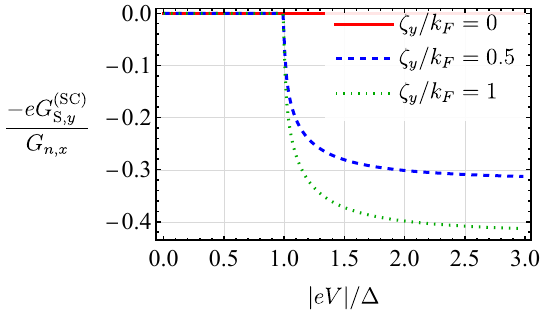}}\end{subfloat}
\begin{subfloat}[]{\includegraphics[width=0.31\columnwidth]{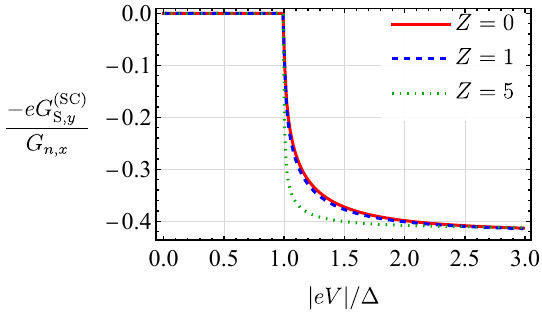}}\end{subfloat}
\caption{\label{fig:sm3-pMSC-G-spin}
The transverse spin conductance $G_{{\rm S}, y}$ for different combinations of parameters as a function of the voltage bias; see Eq.~\eqref{lecm-a-G-S-all} for the definition.
The top and bottom rows show the conductance on the \pmag and SC sides, respectively.
We normalize by the longitudinal normal-state electric conductance $G_{n,x}$ calculated as $G_{{\rm el},x}^{\rm (p{\rm M})}$ at $eV = 10\,\Delta$.
In all panels, unless otherwise stated, $t_{\rm inter}=\mu$, $\zeta_x=0$, $\zeta_y=k_F$, $J=Z=0$, and $\mu=10^3\,\Delta$.
}
\end{figure}

The transverse spin conductance $G_{{\rm S},y}$ is shown in Fig.~\ref{fig:sm3-pMSC-G-spin}. On the \pmag side, the presence of nonzero $\zeta_y$ leads to the transverse spin conductance that, as is explained in the main text, is enhanced by Andreev processes. As with the longitudinal electric conductance, barrier potential suppresses $G_{{\rm S},y}$ for subgap energies leading to a well-pronounced peak at $|eV| \searrow \Delta$, cf. Fig.~\ref{fig:sm3-pMSC-G-el}(c). Unlike the electric conductance, there is a stronger dependence on the spin splitting parameter $\zeta_y$ that determines the spin conductance.

\subsection{NM-SC$p$M junction}
\label{sec:sm3-NMpMSC}

Finally, let us address the transport properties of the NM-SC$p$M junction. 

As with the $p$M-SC junction considered in Sec.~\ref{sec:sm3-pMSC}, we start by presenting the expressions for the longitudinal
\begin{eqnarray}
\label{sm3-NMpMSC-jxElNM}
j_{{\rm el}, x}(x<0) &=& k_{e,+,s} + |a_{s}|^2 k_{h,+, -s} \Theta{\left(-|\IM{k_{h,+,-s}}|\right)} +|b_{s}|^2 k_{e,-,s} \Theta{\left(-|\IM{k_{e,-,s}}|\right)},\\
\label{sm3-NMpMSC-jxElSC}
j_{{\rm el}, x}(x>0) &=& \left(|u|^2 +|v|^2 \right) \left[|c_{s}|^2 \left(p_{e,+,s} +s \zeta_x\right) \Theta{\left(-\left|\IM{p_{e,+,s}}\right|\right)} +|d_{s}|^2 \left(p_{h,-,-s}+s \zeta_x\right)  \Theta{\left(-\left|\IM{p_{h,-,-s}}\right|\right)} \right],\\
\label{sm3-NMpMSC-jxSpinNM}
j_{{\rm S}, x}(x<0) &=& s \left[ k_{e,+,s} - |a_{s}|^2 k_{h,+, -s} \Theta{\left(-|\IM{k_{h,+,-s}}|\right)} +|b_{s}|^2 k_{e,-,s} \Theta{\left(-|\IM{k_{e,-,s}}|\right)}\right], \\
\label{sm3-NMpMSC-jxSpinSC}
j_{{\rm S}, x}(x>0) &=& \frac{s t_{\rm inter}}{\tilde{J}} \left(|u|^2  -|v|^2 \right) \!\left[
\left(p_{e,+,s} + s\zeta_x\right) |c_s|^2 \Theta{\left(-|\IM{p_{e,+,s}}|\right)} -
\left(p_{h,-,-s} + s \zeta_x\right) |d_s|^2 \Theta{\left(-|\IM{p_{h,-,-s}}|\right)} \right] \nonumber\\
\end{eqnarray}
and averaged transverse 
\begin{eqnarray}
\label{sm3-NMpMSC-jyElNM}
\aver{j_{{\rm el}, y}(x<0)} &=& k_y \left[1 + |a_{s}|^2 \Theta{\left(-|\IM{k_{h,+,-s}}|\right)} +|b_{s}|^2 \Theta{\left(-|\IM{k_{e,-,s}}|\right)}\right],\\
\label{sm3-NMpMSC-jyElSC}
\aver{j_{{\rm el}, y}(x>0)} &=& (k_y +s\zeta_y) \left(|u|^2 +|v|^2 \right)\left[|c_{s}|^2 \Theta{\left(-\left|\IM{p_{e,+,s}}\right|\right)} +|d_{s}|^2  \Theta{\left(-\left|\IM{p_{h,-,-s}}\right|\right)}\right],\\
\label{sm3-NMpMSC-jySpinNM}
\aver{j_{{\rm S}, y}(x<0)} &=& s k_y \left[ 1 - |a_{s}|^2 \Theta{\left(-|\IM{k_{h,+,-s}}|\right)}  +|b_{s}|^2 \Theta{\left(-|\IM{k_{e,-,s}}|\right)} \right],\\
\label{sm3-NMpMSC-jySpinpM}
\aver{j_{{\rm S}, y}(x>0)} &=& s \frac{t_{\rm inter}}{\tilde{J}} \left(k_y +s \zeta_y\right) \left(|u|^2  -|v|^2 \right) \left[ |c_s|^2 \Theta{\left(-|\IM{p_{e,+,s}}|\right)} -|d_s|^2 \Theta{\left(-|\IM{p_{h,-,-s}}|\right)} \right]
\end{eqnarray}
currents.

The transport coefficients $a_{s}$, $b_s$, $c_s$, and $d_s$ are given by Eqs.~\eqref{lecm-andreev-2-a}--\eqref{lecm-andreev-2-d} with the replacements $p_{e,+,+} \leftrightarrow k_{e,+,+}$ and $k_{s} \leftrightarrow p_{s}$, where $k_{e,+,+} = \sqrt{2m(\mu+t_{\rm inter}) -k_y^2}$ and $p_{s} = \sqrt{2m(\mu+\tilde{J})+\zeta^2 -(k_y+s\zeta_y)^2}$. 

The transport coefficients on the non-superconducting side $|a_{s}|^2$ and $|b_{s}|^2$ are similar to those in the $p$M-SC junction, see Fig.~\ref{fig:lecm-a-2-coefficients}, with the only quantitative difference being the normal-reflection coefficient $|b_{s}|^2$: it is nonzero for $k_y$ determined by the NM Fermi surface.
Therefore, the transverse electric and longitudinal spin conductances vanish due to symmetry reasons.

\begin{figure}[ht!]
\centering
\begin{subfloat}[]{\includegraphics[width=0.31\columnwidth]{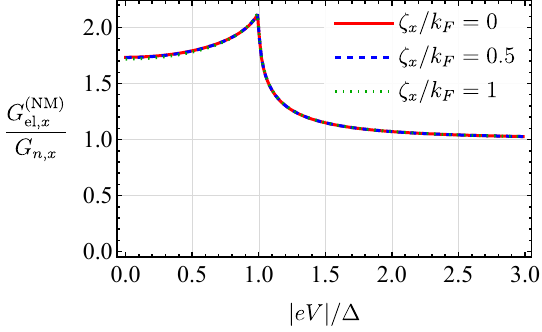}}\end{subfloat}
\begin{subfloat}[]{\includegraphics[width=0.31\columnwidth]{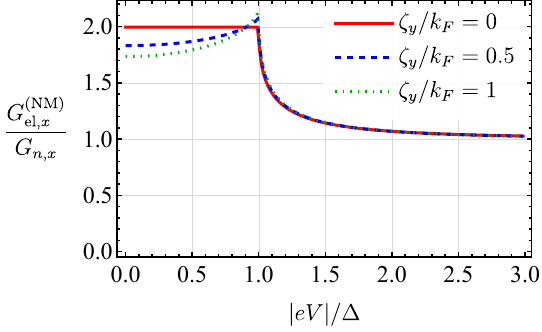}}\end{subfloat}
\begin{subfloat}[]{\includegraphics[width=0.31\columnwidth]{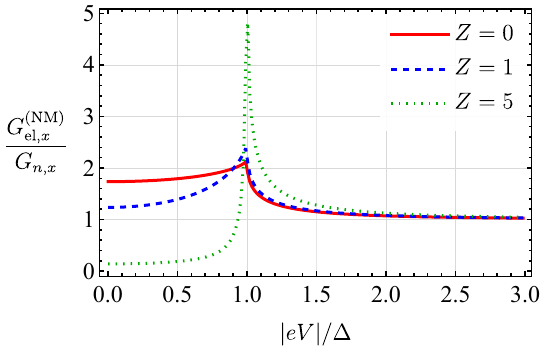}}\end{subfloat}
\\
\begin{subfloat}[]{\includegraphics[width=0.31\columnwidth]{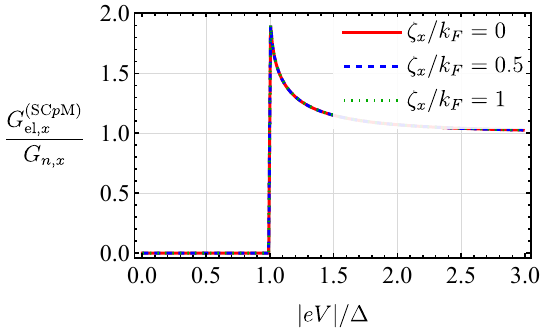}}\end{subfloat}
\begin{subfloat}[]{\includegraphics[width=0.31\columnwidth]{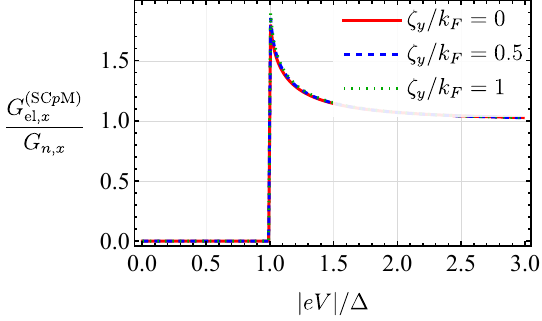}}\end{subfloat}
\begin{subfloat}[]{\includegraphics[width=0.31\columnwidth]{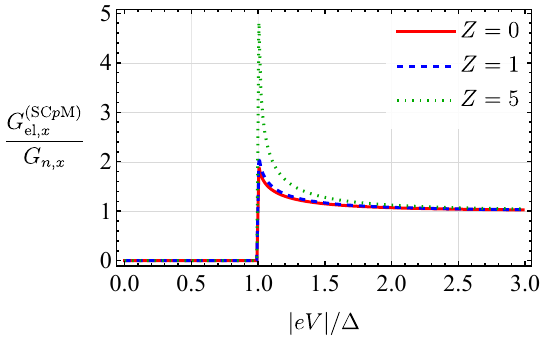}}\end{subfloat}
\caption{\label{fig:sm3-NMpMSC-G-el}
The longitudinal electric conductance $G_{{\rm el},x}$ for different combinations of parameters as a function of the voltage bias; see Eq.~\eqref{lecm-a-G-el-all} for the definition.
The top and bottom rows show the conductance on the NM and SC$p$M sides, respectively.
We normalize by the longitudinal normal-state electric conductance $G_{n,x}$ calculated as $G_{{\rm el},x}^{\rm (NM)}$ at $eV = 10\,\Delta$.
In all panels, unless otherwise stated, $t_{\rm inter}=\mu$, $\zeta_x=0$, $\zeta_y=k_F$, $J=Z=0$, and $\mu=10^3\,\Delta$.
}
\end{figure}

In calculating the longitudinal electric \eqref{lecm-a-G-el-all} and transverse spin \eqref{lecm-a-G-S-all} conductances, we use the expressions in Eqs.~\eqref{sm3-NMpMSC-jxElNM}--\eqref{sm3-NMpMSC-jySpinpM}.
We show the longitudinal electric conductance $G_{{\rm el}, x}$ in Fig.~\ref{fig:sm3-NMpMSC-G-el}. By comparing these results with those in Fig.~\ref{fig:sm3-pMSC-G-el}, we see that $G_{{\rm el}, x}$ is similar in $p$M-SC and NM-SC$p$M junctions. The reason is that the longitudinal electric conductance is affected mostly by the overlap of the Fermi surfaces rather than the details of their spin structure. 

\begin{figure}[ht!]
\centering
\begin{subfloat}[]{\includegraphics[width=0.31\columnwidth]{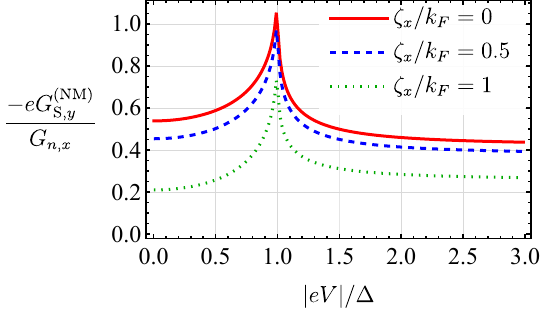}}\end{subfloat}
\begin{subfloat}[]{\includegraphics[width=0.31\columnwidth]{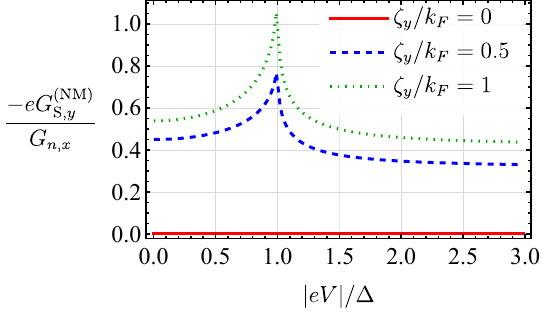}}\end{subfloat}
\begin{subfloat}[]{\includegraphics[width=0.31\columnwidth]{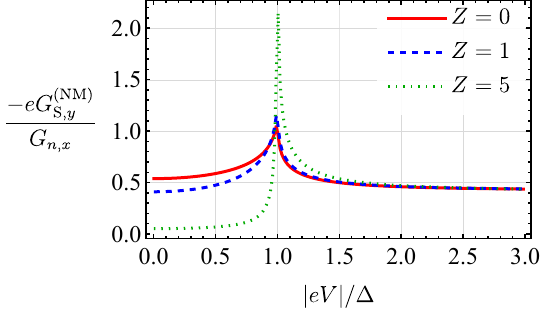}}\end{subfloat}
\\
\begin{subfloat}[]{\includegraphics[width=0.31\columnwidth]{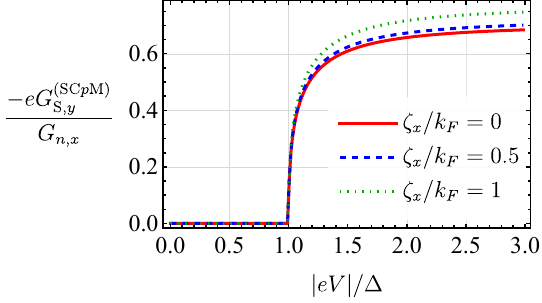}}\end{subfloat}
\begin{subfloat}[]{\includegraphics[width=0.31\columnwidth]{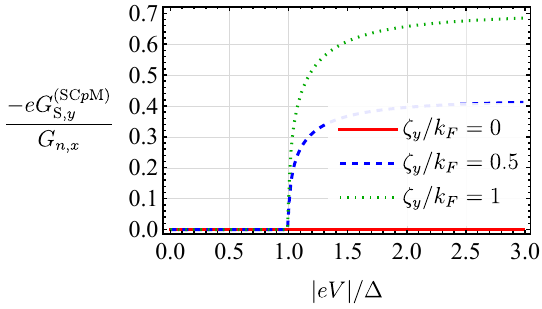}}\end{subfloat}
\begin{subfloat}[]{\includegraphics[width=0.31\columnwidth]{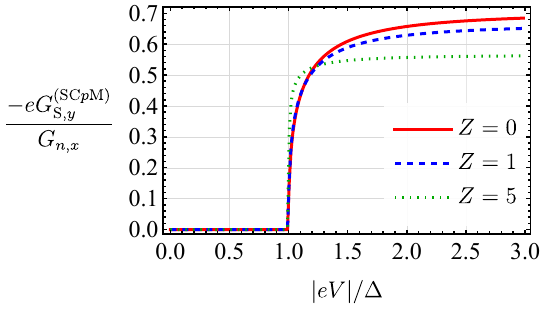}}\end{subfloat}
\caption{\label{fig:sm3-NMpMSC-G-S}
The transverse spin conductance $G_{{\rm S}, y}$ for different combinations of parameters as a function of the voltage bias; see Eq.~\eqref{lecm-a-G-S-all} for the definition.
The top and bottom rows show the conductance on the NM and SC$p$M sides, respectively.
We normalize by the longitudinal normal-state electric conductance $G_{n,x}$ calculated as $G_{{\rm el},x}^{\rm (NM)}$ at $eV = 10\,\Delta$.
In all panels, unless otherwise stated, $t_{\rm inter}=\mu$, $\zeta_x=0$, $\zeta_y=k_F$, $J=Z=0$, and $\mu=10^3\,\Delta$.
}
\end{figure}

The transverse spin conductance $G_{{\rm S}, y}$ is shown in Fig.~\ref{fig:sm3-NMpMSC-G-S}. As with $G_{{\rm el}, x}$, the transverse spin conductance is similar in $p$M-SC and NM-SC$p$M junctions with the only major difference being the sign of the conductance.

